\documentclass[a4paper,11pt]{article}
\pdfoutput=1
\usepackage{jcappub} 
\usepackage[latin9]{inputenc}
\usepackage{textcomp}
\usepackage{amsmath}
\usepackage{amssymb}
\usepackage{color}

\makeatletter

\newcommand{\be}[1]{\begin{equation}\label{#1}}
\newcommand{\ee}{\end{equation}}
\newcommand{\ba}[1]{\begin{eqnarray}\label{#1}}
\newcommand{\ea}{\end{eqnarray}}

\makeatother
\begin{document}

\title{Rotating black holes without $\mathbb{Z}_2$ symmetry and their shadow images}
\author[a,b]{Che-Yu Chen}
\affiliation[a]{Department of Physics and Center for Theoretical Sciences, National Taiwan University, Taipei, Taiwan 10617}
\affiliation[b]{LeCosPA, National Taiwan University, Taipei, Taiwan 10617}

\emailAdd{b97202056@gmail.com}

\abstract{The recent detection of gravitational waves from black hole coalescences and the first image of the black hole shadow enhance the possibilities of testing gravitational theories in the strong-field regime. In this paper, we study the physical properties and the shadow image of a class of Kerr-like rotating black holes, whose $\mathbb{Z}_2$ symmetry is generically broken. Such black hole solutions could arise in effective low-energy theories of a fundamental quantum theory of gravity, such as string theory. Within a theory-agnostic framework, we require that the Kerr-like solutions are asymptotically flat, and assume that a Carter-like constant is preserved, enabling the geodesic equations to be fully separable. Subject to these two requirements, we find that the $\mathbb{Z}_2$ asymmetry of the spacetime is characterized by two arbitrary functions of polar angle. The shadow image turns out to be $\mathbb{Z}_2$ symmetric on the celestial coordinates. Furthermore, the shadow is completely blind to one of the arbitrary functions. The other function, although would affect the apparent size of the shadow, it hardly distorts the shadow contour and has merely no degeneracy with the spin parameter. Therefore, the parameters in this function can be constrained with black hole shadows, only when the mass and the distance of the black hole from the earth are measured with great precision.}

\maketitle
\flushbottom

\section{Introduction}
Recently, with the increasing number of independent observations, it is generally believed that there is a supermassive black hole at the center of galaxies, including the Sagittarius A$^*$ (Sgr A$^*$) in our own Milky Way. One promising way of determining the existence of a black hole is through its shadow.{\footnote{Strictly speaking, the existence of a shadow is just a necessary condition for the existence of a black hole. In fact, any compact object could cast its own shadow as long as the gravitational field at its vicinity is strong enough, such that there is a photon region around the object \cite{Cunha:2017eoe}.}} Imagine there is a background of light sources behind the black hole. When the photons emitted from the light sources pass through the vicinity of the black hole on their way toward the observer, the strong gravitational fields significantly bend the photon trajectories. For the observer, there would appear a dark spot at the line of sight to the black hole because the event horizon would capture those photons whose trajectories are too close to the black hole. This dark spot is the so-called black hole shadow. Essentially, the edge of a shadow is defined by the impact parameter of the photon region around the black hole, on which photons undergo spherical motions. These spherical photon motions within the photon region are unstable against radial perturbations in the sense that any small perturbation on the photon would either make it fall into the black hole or escape to spatial infinity. In Ref.~\cite{Falcke:1999pj}, the authors predicted the capability of observing the shadow of Sgr A$^*$ with very long-baseline interferometry at sub-millimeter wavelengths. With the global collaboration, the shadow of the supermassive black hole at the center of M87 galaxy has been observed \cite{Akiyama:2019cqa,Akiyama:2019brx,Akiyama:2019sww,Akiyama:2019bqs,Akiyama:2019fyp,Akiyama:2019eap}. The first shadow image of Sgr A$^*$ is also expected in the very near future.

Since the observation of black hole shadow is able to reveal the spacetime structure near the black hole, it thus provides us with a potential opportunity to test the nature of the regime with strong gravitational fields. According to Einstein's General Relativity (GR), the rotating black holes in our Universe can be well-described by the Kerr spacetime, which is characterized by only two parameters: the mass and the spin of the black hole. However, there are still some motivations to speculate whether GR is really the complete theory of gravity, in spite of its tremendous success in describing our Universe. One of the reasons is that GR predicts the existence of spacetime singularities, where the theory itself inevitably ceases to be valid. Also, GR is incompatible with quantum mechanics, which means that it would be improper to describe the physics with extremely high energies and small scales in the framework of GR only. In principle, any modifications of GR would change the spacetime structure of Kerr black holes. These deviations from the Kerr spacetime are likely to be significant only inside the event horizon, while remain small at the exterior side. In any case, it is very interesting to see whether these Kerr-like black holes can be tested by observing their shadow images \cite{Cunha:2018acu,Bambi:2019tjh}, investigating the gravitational waves \cite{Yagi:2016jml,Barack:2018yly}, or using other astrophysical tests \cite{Bambi:2015kza}.  

The investigation of black hole shadows has been a very intensive field of research in the last decades, which was preceded by the seminal studies of the shadows of Schwarzschild black holes \cite{Synge:1966okc}, Kerr black holes \cite{Bardeen1973}, and Kerr-Newman black holes \cite{Takahashi:2005hy}. In the literature, the scrutiny following this direction includes, but is not limited to, the shadow of regular black holes \cite{Li:2013jra,Tsukamoto:2014tja,Amir:2016cen,Abdujabbarov:2016hnw,Tsukamoto:2017fxq,Kumar:2019pjp,Jusufi:2019caq,Liu:2020ola}, wormholes \cite{Bambi:2013nla,Nedkova:2013msa,Shaikh:2018kfv,Gyulchev:2018fmd}, black holes coupled with additional fields \cite{Li:2019lsm,Amarilla:2013sj,Wei:2013kza,Cunha:2015yba,Atamurotov:2015xfa,Ali:2019khp,Zhu:2019ura,Ding:2019mal,Ovgun:2018tua,Allahyari:2019jqz,Kumar:2020hgm,Contreras:2019nih,Giddings:2016btb}, black holes with non-commutative geometry \cite{Wei:2015dua,Sharif:2016znp,2018GReGr..50..103S}, black holes in modified theories of gravity \cite{Mizuno:2018lxz,Wang:2018prk,Amarilla:2010zq,Atamurotov:2013dpa,Moffat:2015kva,Cunha:2016wzk,Hennigar:2018hza,Ayzenberg:2018jip,Held:2019xde,Kumar:2019ohr,Konoplya:2020bxa,Wei:2020ght}, black holes in higher dimensions \cite{Vagnozzi:2019apd,Amarilla:2011fx,Eiroa:2017uuq,Long:2019nox,Banerjee:2019nnj}, and black holes surrounded by dark matter \cite{Hou:2018avu,Haroon:2018ryd,Konoplya:2019sns,2019arXiv190511803J}. Furthermore, the effects from cosmic expansion on black hole shadows have recently been investigated \cite{Perlick:2018iye,Bisnovatyi-Kogan:2018vxl,Tsupko:2019pzg,Qi:2019zdk,Tsupko:2019mfo,Vagnozzi:2020quf,Li:2020drn}. In addition to studying black holes of some specific models, one could consider testing black hole solutions from a theory-agnostic perspective. More precisely, one could construct a Kerr-like spacetime which is characterized by some deviation functions. These functions and their parameters essentially parametrize possible deviations of the spacetime from its Kerr counterpart. Although such parametrized Kerr-like black holes may not be solutions to any particular gravitational theory, constraining the deviation functions would enable us to test the Kerr spacetime, as well as the no-hair theorem. Such a \textit{post-Kerr} approach has been widely adopted to test Kerr spacetime by using astrophysical observations. There have been various parametrized Kerr-like spacetimes in the literature \cite{Vigeland:2011ji,Glampedakis:2005cf,Johannsen:2011dh,Cardoso:2014rha,Johannsen:2015pca,Ghasemi-Nodehi:2016wao,Cardoso:2015xtj,Konoplya:2016jvv,Konoplya:2018arm,Konoplya:2016pmh,Carson:2020dez}. For some scrutinies of these parametrized Kerr-like black holes and their shadows, we refer the readers to Refs.~\cite{Ghasemi-Nodehi:2015raa,Johannsen:2015qca,Younsi:2016azx,Wang:2017hjl,Abdikamalov:2019ztb,Shaikh:2019fpu}.

In this paper, we will study the shadow of a class of Kerr-like black holes, in which the deviation functions generically break the $\mathbb{Z}_2$ symmetry of the spacetime. In the literature, we find that less attention has been paid to this particular class of Kerr-like black holes. However, black hole solutions whose $\mathbb{Z}_2$ symmetry is broken would generically appear in gravitational theories containing parity-violating terms. In the effective field theory approach toward quantum gravity, one considers a series of correction terms on top of the Einstein-Hilbert action. In general, these correction terms contain higher order terms of curvature which, at the level of field equations, appear in the form of higher-derivative interactions. The parity-violating corrections are generally possible and they are characterized by higher order curvature terms consisting of the dual Riemann tensor:
\begin{equation}
\tilde R_{\mu\nu\alpha\beta}\equiv\frac{1}{2}\epsilon_{\mu\nu\rho\sigma}{R^{\rho\sigma}}_{\alpha\beta}\,.
\end{equation}
In Ref.~\cite{Cano:2019ore}, it has been shown that when the Chern-Simons term is coupled with the Gauss-Bonnet term through the dynamical Chern-Simons scalar field, the $\mathbb{Z}_2$ symmetry of a rotating black hole solution would be broken. In addition, such a $\mathbb{Z}_2$ symmetry violation is also found in the effective field theory containing higher order of curvature invariants constructed by $\tilde R_{\mu\nu\alpha\beta}$ \cite{Cardoso:2018ptl}. 

In this paper, within a theory-agnostic framework and starting with a general metric given in Ref.~\cite{Papadopoulos:2018nvd}, we will construct a set of Kerr-like black holes, whose deviation functions generically break the $\mathbb{Z}_2$ symmetry of the spacetime. We will require that the spacetime reduces to the Kerr solution at a far distance from the black hole, and that there exists a Carter-like constant in the spacetime. As in the Kerr spacetime in which the existence of a Carter constant corresponds to a hidden symmetry characterized by the Killing tensor, we expect that this property of the Kerr-like spacetime is preserved. In addition, the existence of a Carter-like constant allows for the full separability of the geodesic equations. We will show that the Kerr-like spacetime is characterized by two arbitrary functions of polar angle, which generically break the $\mathbb{Z}_2$ symmetry of the spacetime. It turns out that the shadow contour is completely blind to one of the deviation functions. Also, the other deviation function seems not to alter the overall shape of the shadow, but it does change the apparent size of the silhouette.

This paper is outlined as follows. Before formulating the Kerr-like solutions, in section~\ref{secII} we briefly review the general axisymmetric metric proposed in Ref.~\cite{Papadopoulos:2018nvd} and then derive its null geodesic equations. In section~\ref{sec.Kerrme}, we construct a class of Kerr-like black hole metrics, focusing on the deviations that break the $\mathbb{Z}_2$ symmetry of the spacetime. We will also exhibit how the $\mathbb{Z}_2$ asymmetry appears near the horizon and the ergosurface. In section~\ref{sec.shadow}, we investigate the shadow contour of the Kerr-like black hole and see how the deviation function change the size and the shape of the shadow. We finally present our conclusions in section~\ref{sec.conclu}.

\section{The PK metric and the null geodesic equations}\label{secII}
As we have mentioned in the Introduction, we will investigate the shadow of a class of Kerr-like black holes, whose $\mathbb{Z}_2$ symmetry is generically broken. In the construction of such a class of spacetimes, we will respect one of the important symmetries in the original Kerr spacetime: the existence of a Carter-like constant \cite{Carter:1968ks,BenentiFrancaviglia}. Subject to this assumption, the null geodesic equations are then completely separable. In Ref.~\cite{Papadopoulos:2018nvd}, Papadopoulos and Kokkotas developed an innovative approach to construct the most general axisymmetric spacetime (the PK metric) which preserves the Carter-like constant and the separability of the geodesic equations. In the theory-agnostic framework, the PK metric turns out to be a suitable metric to parametrize the possible deviations in the Kerr-like spacetime of our interest. 

In the Boyer-Lindquist coordinate system $(t,r,y,\psi)$ where $y\equiv\cos\theta$, the contravariant form of the PK metric can be written as \cite{Papadopoulos:2018nvd}
\begin{align}
g^{tt}&=\frac{\mathcal{A}_5(r)+\mathcal{B}_5(y)}{\mathcal{A}_1(r)+\mathcal{B}_1(y)}\,,\quad g^{t\psi}=\frac{\mathcal{A}_4(r)+\mathcal{B}_4(y)}{\mathcal{A}_1(r)+\mathcal{B}_1(y)}\,,\nonumber\\
g^{\psi\psi}&=\frac{\mathcal{A}_3(r)+\mathcal{B}_3(y)}{\mathcal{A}_1(r)+\mathcal{B}_1(y)}\,,\quad g^{yy}=\frac{\mathcal{B}_2(y)}{\mathcal{A}_1(r)+\mathcal{B}_1(y)}\,,\nonumber\\
g^{rr}&=\frac{\mathcal{A}_2(r)}{\mathcal{A}_1(r)+\mathcal{B}_1(y)}\,.\label{PK}
\end{align}
In the above expression, $\mathcal{A}_i(r)$ and $\mathcal{B}_i(y)$ are arbitrary functions of $r$ and $y$, respectively. It has been proven that the PK metric \eqref{PK} allows for the existence of a Cater-like constant and the null geodesic equations as well as the corresponding Hamilton-Jacobi equation are fully separable. In the following discussion, we will exhibit the separability of the null geodesic equations of this general spacetime. It should be emphasized that in the literature, several parametrized Kerr-like metrics \cite{Vigeland:2011ji,Johannsen:2015pca,Konoplya:2018arm,Carson:2020dez}, whose geodesic equations are required to be separable, are actually subclasses of the PK metric. The relation among some of these parametrized metrics is discussed in Ref.~\cite{Chen:2019jbs}.

\subsection{Null geodesic equations}
In this subsection, we will derive the null geodesic equations of the general PK metric \eqref{PK} and exhibit their separability by using the Hamilton-Jacobi approach. The set of geodesic equations of the metric \eqref{PK} is described by the following Lagrangian:
\begin{equation}
\mathcal{L}=\frac{1}{2}g_{\mu\nu}\dot{x}^\mu\dot{x}^\nu=\frac{1}{2}\left(g_{tt}\dot{t}^2+g_{rr}\dot{r}^2+g_{yy}\dot{y}^2+g_{\psi\psi}\dot{\psi}^2+2g_{t\psi}\dot{t}\dot{\psi}\right)\,,\label{lagrangianL}
\end{equation}
where the dot denotes the derivative with respect to an affine parameter $\tau$. It can be immediately seen from the Lagrangian \eqref{lagrangianL} that there are two constants of motion in the system: the conserved energy $E\equiv-\partial\mathcal{L}/\partial \dot{t}$ and the conserved azimuthal angular momentum $L_z\equiv\partial\mathcal{L}/\partial\dot{\psi}$. Using these conserved quantities, one gets the following two geodesic equations
\begin{align}
\dot{t}&=\frac{Eg_{\psi\psi}+L_zg_{t\psi}}{g_{t\psi}^2-g_{tt}g_{\psi\psi}}\,,\label{dott}\\
\dot{\psi}&=-\frac{Eg_{t\psi}+L_zg_{tt}}{g_{t\psi}^2-g_{tt}g_{\psi\psi}}\,.\label{dotpsi}
\end{align} 
Using the PK metric coefficients given in Eqs.~\eqref{PK}, the geodesic equations \eqref{dott} and \eqref{dotpsi} can be rewritten as
\begin{align}
\left(\mathcal{A}_1+\mathcal{B}_1\right)\dot{t}&=-E\left(\mathcal{A}_5+\mathcal{B}_5\right)+L_z\left(\mathcal{A}_4+\mathcal{B}_4\right)\,,\label{dott2}\\
\left(\mathcal{A}_1+\mathcal{B}_1\right)\dot{\psi}&=-E\left(\mathcal{A}_4+\mathcal{B}_4\right)+L_z\left(\mathcal{A}_3+\mathcal{B}_3\right)\,,\label{dotpsi2}
\end{align}
respectively.

Then, we consider the geodesic equations of $r$ and $y$. In the general PK spacetime, there exists a Carter-like constant and this ensures the separability of these two geodesic equations. The geodesic equations for $r$ and $y$ can be derived by considering the Hamilton-Jacobi equation
\begin{equation}
\frac{\partial\mathcal{S}}{\partial\tau}+\mathcal{H}=0\,,\label{HJeq1}
\end{equation}
where $\mathcal{S}$ is the Jacobi action and $\mathcal{H}$ is the Hamiltonian. The Hamiltonian associated with the Lagrangian \eqref{lagrangianL} can be written as  
\begin{equation}
\mathcal{H}=\frac{1}{2}p_\mu p^\mu\,,
\end{equation}
where $p_\mu$ is the conjugate momentum and it can be expressed as follows
\begin{equation}
p_\mu\equiv\frac{\partial\mathcal{L}}{\partial\dot{x}^\mu}=g_{\mu\nu}\dot{x}^\nu=\frac{\partial\mathcal{S}}{\partial x^\mu}\,.\label{pmuexpre}
\end{equation}
Therefore, the Hamilton-Jacobi equation \eqref{HJeq1} can be written as
\begin{equation}
\frac{\partial\mathcal{S}}{\partial\tau}=-\frac{1}{2}g^{\mu\nu}\frac{\partial\mathcal{S}}{\partial x^\mu}\frac{\partial\mathcal{S}}{\partial x^\nu}\,.\label{HJeq2}
\end{equation}

For a separable Hamilton-Jacobi equation, we can write the Jacobi action as follows
\begin{equation}
\mathcal{S}=\frac{1}{2}\epsilon\tau-Et+L_z\psi+\mathcal{S}_r(r)+\mathcal{S}_y(y)\,,\label{jacobiactionse}
\end{equation}
where $\epsilon=0$ for photon geodesics. Note that the geodesic equations for massive particles ($\epsilon=1$) are also separable. Inserting the ansatz \eqref{jacobiactionse} into the Hamilton-Jacobi equation \eqref{HJeq2}, one obtains
\begin{equation}
\mathcal{A}_5E^2-2\mathcal{A}_4EL_z+\mathcal{A}_3L_z^2+\mathcal{A}_2\left(\frac{dS_r}{dr}\right)^2=-\mathcal{B}_5E^2+2\mathcal{B}_4EL_z-\mathcal{B}_3L_z^2-\mathcal{B}_2\left(\frac{dS_y}{dy}\right)^2\,.\label{sepeq}
\end{equation}
It can be seen that the left-hand side of Eq.~\eqref{sepeq} only depends on $r$, while the right-hand side only depends on $y$. Therefore, this equation is separable by introducing a decoupling constant (the Carter-like constant $K$) and we obtain:

\begin{align}
\left(\mathcal{A}_1+\mathcal{B}_1\right)\dot{r}&=\pm\sqrt{R(r)}\,,\label{dotr}\\
\left(\mathcal{A}_1+\mathcal{B}_1\right)\dot{\theta}&=\pm\sqrt{\Theta(\theta)}\,,\label{dottheta}
\end{align}
where
\begin{align}
R(r)&\equiv \mathcal{A}_2\left[-\mathcal{A}_5E^2+2\mathcal{A}_4EL_z-\mathcal{A}_3L_z^2-K-\left(L_z-aE\right)^2\right]\,,\\
\Theta(\theta)&\equiv \left[-\mathcal{B}_5E^2+2\mathcal{B}_4EL_z-\mathcal{B}_3L_z^2+K+\left(L_z-aE\right)^2\right]\left(\frac{\mathcal{B}_2}{\sin^2\theta}\right)\,,
\end{align}
where $a$ stands for the spin of the spacetime. Note that $\mathcal{B}_i$ are functions of $y$, which are also functions of polar angle $\theta$ via $y=\cos\theta$. In the derivation of Eqs.~\eqref{dotr} and \eqref{dottheta}, we have used the last equality in Eq.~\eqref{pmuexpre}. 

For the sake of later convenience, we define the following parameters: $\xi\equiv L_z/E$ and $\eta\equiv K/E^2$, such that 
\begin{align}
\frac{R(r)}{E^2}&=\mathcal{A}_2\left[-\mathcal{A}_5+2\mathcal{A}_4\xi-\mathcal{A}_3\xi^2-\eta-\left(\xi-a\right)^2\right]\,,\label{RE}\\
\frac{\Theta(\theta)}{E^2}&=\left[-\mathcal{B}_5+2\mathcal{B}_4\xi-\mathcal{B}_3\xi^2+\eta+\left(\xi-a\right)^2\right]\left(\frac{\mathcal{B}_2}{\sin^2\theta}\right)\,.\label{ThetaE}
\end{align}
The geodesic equations for $\dot{t}$ and $\dot{\psi}$, i.e., Eqs.~\eqref{dott2} and \eqref{dotpsi2}, can then be written as 
\begin{align}
\left(\mathcal{A}_1+\mathcal{B}_1\right)\dot{t}/E&=-\left(\mathcal{A}_5+\mathcal{B}_5\right)+\xi\left(\mathcal{A}_4+\mathcal{B}_4\right)\,,\label{dott3}\\
\left(\mathcal{A}_1+\mathcal{B}_1\right)\dot{\psi}/E&=-\left(\mathcal{A}_4+\mathcal{B}_4\right)+\xi\left(\mathcal{A}_3+\mathcal{B}_3\right)\,.\label{dotpsi3}
\end{align}
It should be highlighted that Eqs.~\eqref{dotr}, \eqref{dottheta}, \eqref{dott3}, and \eqref{dotpsi3}, with the functions $R(r)$ and $\Theta(\theta)$ given in Eqs.~\eqref{RE} and \eqref{ThetaE}, stand for the null geodesic equations of the general PK metric \eqref{PK}. They are completely separable and can be written in first-order form. Also, as we have mentioned, the geodesic equations for massive particles ($\epsilon=1$) are still fully separable. Since in this paper, we will investigate the shadow contour generated by photons, we will only focus on the photon geodesics ($\epsilon=0$).

\section{Kerr-like black holes without $\mathbb{Z}_2$ symmetry}\label{sec.Kerrme}
The PK metric \eqref{PK} is described by five arbitrary functions of $r$ and five arbitrary functions of $y$, that is $\mathcal{A}_i(r)$ and $\mathcal{B}_i(y)$, respectively. In order to focus on the Kerr-like black holes of our interest, these arbitrary functions should be fixed to some extent such that they can properly parametrize how the spacetime deviates from the Kerr geometry. As we have mentioned in the Introduction, the formulation of black hole solutions within the effective field theories to extend GR \cite{Cardoso:2018ptl,Cano:2019ore}, or in the generic presence of non-minimal matter couplings \cite{Cunha:2018uzc}, strongly motivate the consideration of black hole spacetimes without $\mathbb{Z}_2$ symmetry. In this paper, we study the physical properties of such black hole spacetimes in a theory-agnostic framework. More precisely, we construct the Kerr-like black hole metric by assigning properly the functions $\mathcal{A}_i(r)$ and $\mathcal{B}_i(y)$. In particular, we will focus on those deviations that break the $\mathbb{Z}_2$ symmetry of the original Kerr spacetime. Such deviations are encoded in the functions $\mathcal{B}_i(y)$. In principle, if functions $\mathcal{B}_i(y)$ are not even under the parity change, for example, if they contain odd powers of $y$, the spacetime structure would not be invariant under $y\leftrightarrow-y$ exchange and the $\mathbb{Z}_2$ symmetry is broken.  

For the sake of abbreviation, we define $\Delta\equiv r^2-2M(r)r+a^2$ and $X\equiv r^2+a^2$. Then, we construct a Kerr-like black hole by assuming the PK metric functions to be
\begin{align}
\mathcal{A}_1=r^2\,,\quad\mathcal{A}_2=\Delta\,,\quad\mathcal{A}_3=-\frac{a^2}{\Delta}\,,\nonumber\\
\mathcal{A}_4=-\frac{aX}{\Delta}\,,\qquad\mathcal{A}_5=-\frac{X^2}{\Delta}\,,\label{Ametric}
\end{align}
and
\begin{align}
\mathcal{B}_1=a^2y^2+\tilde\epsilon_1(y)\,,\quad\mathcal{B}_2=1-y^2+\tilde\epsilon_2(y)\,,\quad\mathcal{B}_3=\frac{1}{1-y^2}+\tilde\epsilon_3(y)\,,\nonumber\\
\mathcal{B}_4=a+\tilde\epsilon_4(y)\,,\qquad\mathcal{B}_5=a^2(1-y^2)+\tilde\epsilon_5(y)\,,\label{Bmetric}
\end{align}
where $\tilde\epsilon_i(y)$ quantify the deviations from Kerr spacetime in terms of the polar angle $\theta$. Note that we keep the radial dependence in the mass function $M(r)$. In the absence of $\tilde\epsilon_i(y)$ and when $M(r)=M$, the Kerr spacetime is recovered.

Another important requirement for a valid isolated black hole spacetime is asymptotic flatness. We require that the Kerr-like spacetime should reduce to Kerr spacetime when $r\rightarrow\infty$. To implement this, we assume $M(\infty)\rightarrow M$ and interpret $M$ as the mass of the black hole. In the asymptotic region, we find that 
\begin{align}
&g_{tt}=-1+\frac{2M}{r}+\mathcal{O}\left(r^{-2}\right)\,,\qquad g_{rr}=1+\frac{2M}{r}+\mathcal{O}\left(r^{-2}\right)\,,\nonumber\\
&g_{yy}=r^2\left[\frac{1}{1-y^2+\tilde\epsilon_2}+\mathcal{O}\left(r^{-2}\right)\right]\,,\quad g_{\psi\psi}=r^2\left[\frac{1-y^2}{1+\tilde\epsilon_3-y^2\tilde\epsilon_3}+\mathcal{O}\left(r^{-2}\right)\right]\,,\nonumber\\
&g_{t\psi}=\frac{\tilde\epsilon_4}{\frac{1}{1-y^2}+\tilde\epsilon_3}-\frac{2M\left(1-y^2\right)\left(a+\tilde\epsilon_4\right)}{r\left[1+\left(1-y^2\right)\tilde\epsilon_3\right]}+\mathcal{O}\left(r^{-2}\right)\,.
\end{align}
It turns out that one can redefine the coordinate $y$ such that $\tilde\epsilon_2=0$. Also, we have to assume $\tilde\epsilon_3=\tilde\epsilon_4=0$ to respect the asymptotic flatness of the spacetime. After taking these conditions into account, the Kerr-like spacetime is characterized by $\tilde\epsilon_1(y)$, $\tilde\epsilon_5(y)$, and the mass function $M(r)$.{\footnote{In fact, if the mass function is a constant ($M(r)=M$), the coefficient of the $1/r^2$ term in the expansion of $g_{tt}$ is proportional to $\tilde\epsilon_1+\tilde\epsilon_5$. Therefore, the Solar System tests \cite{Williams:2004qba} would give a further constraint: $\tilde\epsilon_1+\tilde\epsilon_5\approx0$. Since we have kept a general radial dependence in the mass function, for the time being we will keep $\tilde\epsilon_1(y)$ and $\tilde\epsilon_5(y)$ as two independent functions for the sake of generality.}}

Before proceeding further, we would like to mention that the $\mathbb{Z}_2$ asymmetry also exists for Kerr-NUT black holes \cite{Newman:1963yy}, in which the symmetry is broken due to a non-vanishing gravitomagnetic charge $l$. The geodesic equations of the Kerr-NUT spacetime are separable as well \cite{Abdujabbarov:2012bn}. Furthermore, the geodesic equations are also separable for the Kerr-Newman-NUT black hole \cite{Mukherjee:2018dmm} and its generalization with cosmological constant \cite{Grenzebach:2014fha}. However, this class of spacetimes is not asymptotically flat in general. In addition, the metric has singularity on the axis of symmetry ($y=\pm1$). This can be seen by mapping the Kerr-NUT metric into the PK metric. We find that
\begin{align}
\mathcal{B}_1&=a^2y^2+2lay\,,\quad \mathcal{B}_4=a-\frac{2ly}{1-y^2}\,,\nonumber\\
\mathcal{B}_5&=a^2\left(1-y^2\right)-4aly+\frac{4l^2y^2}{1-y^2}\,,
\end{align} 
for the Kerr-NUT black hole. In the presence of terms linear in $y$, the $\mathbb{Z}_2$ symmetry is broken. The singularity on the axis of symmetry is due to the $1-y^2$ factors appearing in the denominator of $\mathcal{B}_4$ and $\mathcal{B}_5$. In addition, the spacetime is not asymptotically flat due to the second term of $\mathcal{B}_4$, which corresponds to a non-vanishing $\tilde\epsilon_4$. For the Kerr-like metric given in Eqs.~\eqref{Ametric} and \eqref{Bmetric}, we require $\tilde\epsilon_4=0$. Also, the singularity on the axis of symmetry can be easily avoided by choosing $\tilde\epsilon_1$ and $\tilde\epsilon_5$ properly.

\subsection{Horizon}
Due to the presence of the arbitrary functions $\tilde\epsilon_1(y)$ and $\tilde\epsilon_5(y)$, the black hole spacetime under consideration is generically not $\mathbb{Z}_2$ symmetric. This can happen whenever $\tilde\epsilon_1(y)$ or $\tilde\epsilon_5(y)$ is not even under $y\leftrightarrow-y$ exchange. Such $\mathbb{Z}_2$ asymmetry in the spacetime structure can be visualized near the event horizon and the ergosurface. In this subsection, we first elucidate the spacetime structure on the event horizon in more details.

The event horizon $r_h$ of the Kerr-like black hole is defined by the surface $r=r_h$ such that $\Delta(r_h)=0$.{\footnote{There can be multiple roots for the equation $\Delta=0$. We shall regard the outermost one as the event horizon in the following discussions.}} On this surface, we define the mass function as $M(r_h)\equiv M_h$. In order to prove that $r=r_h$ is indeed the event horizon, we follow the procedure in Ref.~\cite{Cano:2019ore} and first consider the determinant of the $(t,\psi)$ metric:
\begin{equation}
g_{tt}g_{\psi\psi}-g_{t\psi}^2=-\frac{\left(1-y^2\right)\left[r^2+y^2a^2+\tilde\epsilon_1(y)\right]^2\Delta(r)}{\left(r^2+y^2a^2\right)^2-\left(r^2-2M(r)r+y^2a^2\right)\tilde\epsilon_5(y)}\,.
\end{equation}
It can be seen that the determinant vanishes when $\Delta=0$, indicating that the corresponding surface $r=r_h$ is null-like. In addition, the surface $r=r_h$ is a Killing horizon since on this surface there exists a Killing vector $\zeta\equiv\partial_t+\Omega\partial_\psi$, where
\begin{equation}
\Omega\equiv\frac{|g_{t\psi}|}{g_{\psi\psi}}\Bigg|_{r=r_h}=\frac{a}{2M_hr_h}\,,
\end{equation}
is a constant, such that the norm of $\zeta$ vanishes. As a result, the surface $r=r_h$ is indeed the event horizon of the black hole.

Since the most important features of the Kerr-like black hole considered in this paper is the violation of its $\mathbb{Z}_2$ symmetry, it could be interesting to visualize it and see how the spacetime structure is modified when changing $\tilde\epsilon_1(y)$ and $\tilde\epsilon_5(y)$. In order to visualize the horizon structure, we use an isometric embedding to map the horizon geometry into a 3-dimensional Euclidean space. The induced metric on the horizon is
\begin{equation}
ds_h^2=\frac{r_h^2+y^2a^2+\tilde\epsilon_1(y)}{1-y^2}dy^2+\frac{4M_h^2r_h^2\left(1-y^2\right)\left[r_h^2+y^2a^2+\tilde\epsilon_1(y)\right]}{\left(r_h^2+y^2a^2\right)^2+a^2\left(1-y^2\right)\tilde\epsilon_5(y)}d\psi^2\,.\label{3dembedinghorizon}
\end{equation}
The standard procedure \cite{Smarr:1973zz} of the embedding starts with a coordinate mapping from $\left(y,\psi\right)\rightarrow\left(x^1,x^2,x^3\right)$ by
\begin{equation}
x^1=F(y)\cos\psi\,,x^2=F(y)\sin\psi\,,x^3=G(y)\,.
\end{equation}
The resulting 2-metric reads
\begin{equation}
ds^2=d(x^1)^2+d(x^2)^2+d(x^3)^2=\left(F_{,y}^2+G_{,y}^2\right)dy^2+F^2d\psi^2\,.\label{3dembeding}
\end{equation}
Equating Eq~\eqref{3dembedinghorizon} to Eq.~\eqref{3dembeding}, we get
\begin{align}
F(y)&=2M_hr_h\left\{\frac{\left(1-y^2\right)\left[r_h^2+y^2a^2+\tilde\epsilon_1(y)\right]}{\left(r_h^2+y^2a^2\right)^2+a^2\left(1-y^2\right)\tilde\epsilon_5(y)}\right\}^{1/2}\,,\\
G(y)&=\int\left[\frac{r_h^2+y^2a^2+\tilde\epsilon_1(y)}{1-y^2}-F_{,y}^2\right]^{1/2}dy\,.
\end{align}

\begin{figure}
\begin{center}
\includegraphics[scale=0.45]{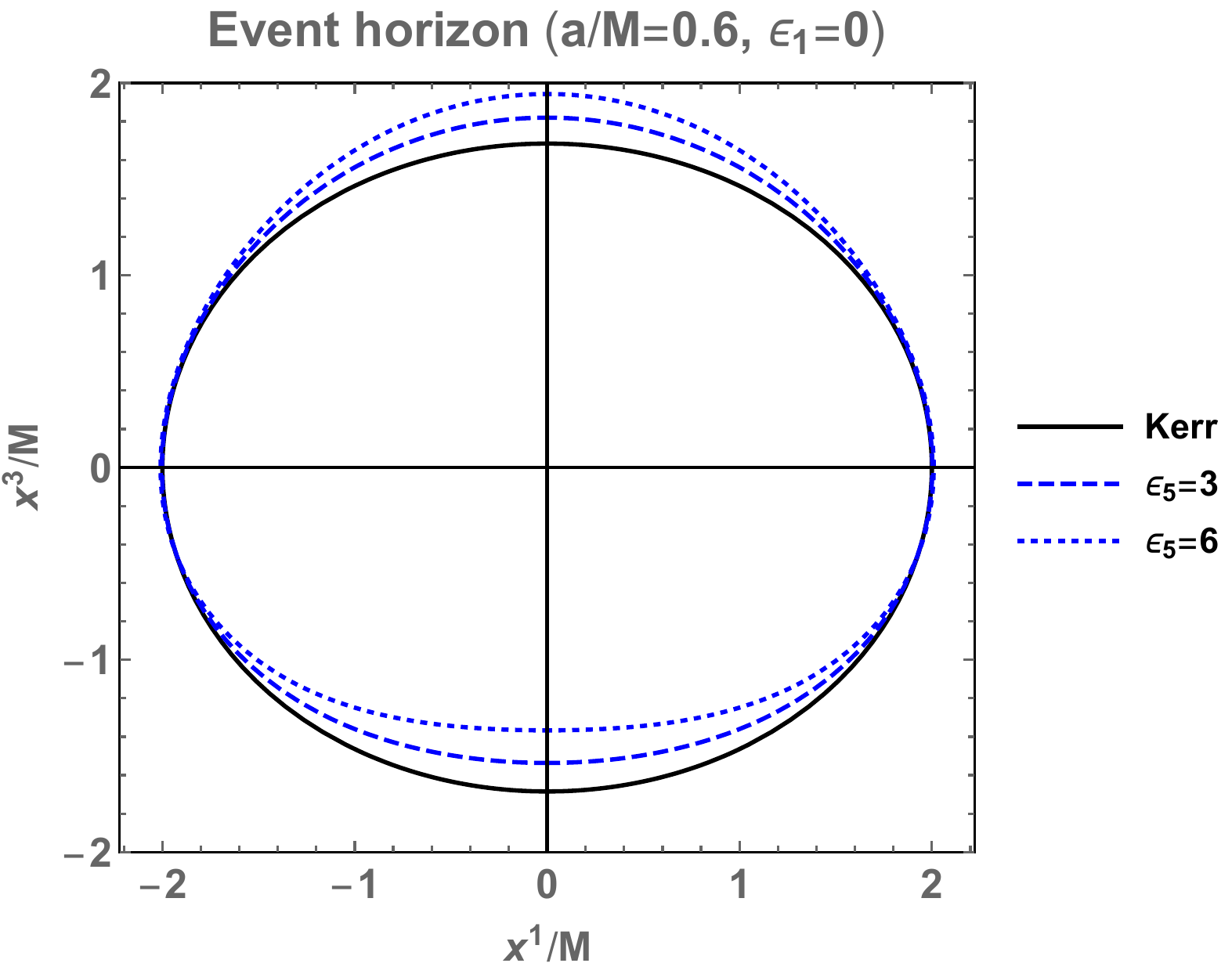}
\includegraphics[scale=0.45]{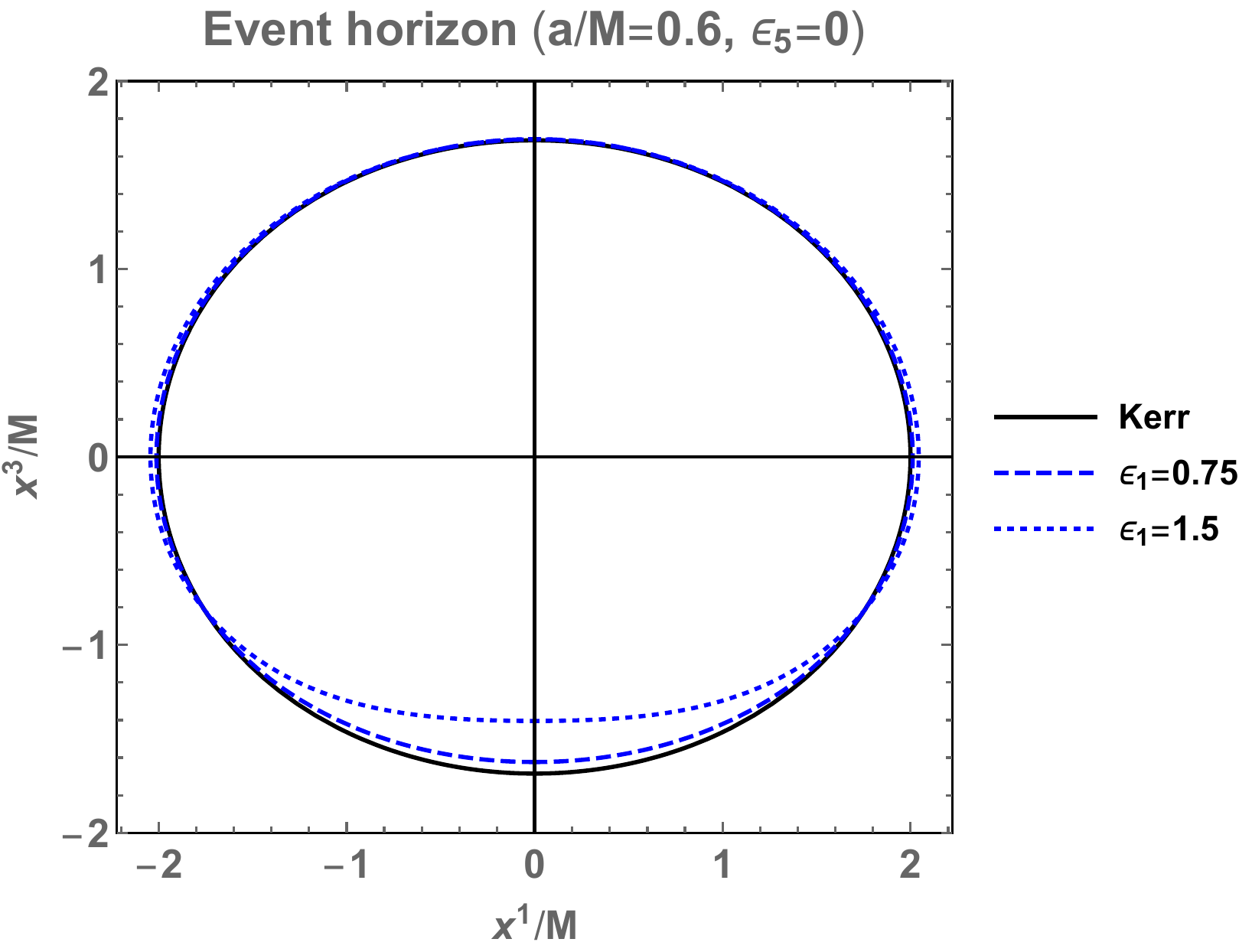}
\caption{\label{fig.embedeh}The embeddings of the event horizon in 3-dimensional Euclidean space. The black contour is the Kerr solution. In the left panel, we fix $\epsilon_1=0$ and $a/M=0.6$, then show the contour for different values of $\epsilon_5$. In the right panel, on the other hand, we fix $\epsilon_5=0$ and $a/M=0.6$, then show the contour for different values of $\epsilon_1$. In both cases, the $\mathbb{Z}_2$ asymmetry can be easily seen when changing $\epsilon_1$ or $\epsilon_5$.}
\end{center}
\end{figure}

In Figure~\ref{fig.embedeh}, we assume $M(r)=M$, $\tilde\epsilon_1(y)=\epsilon_1M^2y$, and $\tilde\epsilon_5(y)=\epsilon_5M^2y$, where $\epsilon_1$ and $\epsilon_5$ are dimensionless constants. The embeddings of the event horizon in 3-dimensional Euclidean space are shown. In the left panel, we fix $\epsilon_1=0$ and $a/M=0.6$, and show the embeddings for different values of $\epsilon_5$. In the right panel, we fix $\epsilon_5=0$ and again $a/M=0.6$, then show the results for different values of $\epsilon_1$. The black contours in both panels correspond to the event horizon of the Kerr black hole. In addition, for each contour in the figure, we have shifted the contour vertically such that the origin of the vertical axis is located at the value of $\theta_m$ which maximizes the proper length of a constant-$\theta $ circle on the horizon. When increasing $\epsilon_5$ from zero, one can see from the left panel of Figure~\ref{fig.embedeh} that both the upper and lower poles are shifted upward. This gives rise to asymmetry of the contour between the upper and lower half-planes. The $\mathbb{Z}_2$ symmetry is therefore broken. In the right panel, on the other hand, one can see that when increasing $\epsilon_1$ from zero, the upper pole of the contour remains almost intact, while the lower pole is shifted upward by a comparable amount. The violation of $\mathbb{Z}_2$ symmetry in this case is more transparent. It should be emphasized that the consideration of a negative value of $\epsilon_1$ or $\epsilon_5$ induces the same shifts of the poles as what are induced for positive values of them, while toward the opposite direction. As a consequence, the non-vanishing functions $\tilde\epsilon_1(y)$ or $\tilde\epsilon_5(y)$ could break the $\mathbb{Z}_2$ symmetry of the spacetime.

\subsection{Ergosurface}
Another crucial surface of the Kerr-like black holes is the ergosurface. This surface defines the boundary of the so-called ergosphere in which any physical object in this region is unlikely to appear stationary with respect to an observer far away from the black hole. Basically, given a black hole spacetime metric in the Boyer-Lindquist coordinate system, the ergosurface $r=r_e$ is defined by $g_{tt}=0$. Using the Kerr-like metric considered in this paper, that is, Eqs.~\eqref{PK} and \eqref{Ametric}, the ergosurface is defined by
\begin{equation}
\textrm{ergosurface:}\qquad r_{e}^2-2M_er_e+a^2y^2=0\,,
\end{equation}
where we have defined $M(r_{e})\equiv M_e$. Similar to the definition of an event horizon, the above equation may contain multiple roots. We shall regard the outermost one as the ergosurface $r_{e}$. The induced metric at the ergosurface is
\begin{align}
ds_{e}^2
=&\,\left[r_{e}^2+y^2a^2+\tilde\epsilon_1(y)\right]\left[\frac{1}{1-y^2}+\frac{1}{\Delta(r_{e})}\left(\frac{dr_{e}}{dy}\right)^2\right]dy^2+\nonumber\\&\frac{\left(1-y^2\right)\left[r_{e}^2+y^2a^2+\tilde\epsilon_1(y)\right]\left\{r_{e}^4+2M_ea^2r_{e}+r_{e}^2a^2+\Delta(r_{e})\left[y^2a^2-\tilde\epsilon_5(y)\right]\right\}}{\left(r_{e}^2+y^2a^2\right)^2}d\psi^2\,\nonumber\\
=&\,\frac{\left[2M_er_{e}+\tilde\epsilon_1(y)\right]M_e^2}{\left[M_e^2-a^2y^2\right]\left(1-y^2\right)}dy^2+\nonumber\\&\frac{\left(1-y^2\right)\left[2M_er_{e}+\tilde\epsilon_1(y)\right]\left[4M_e^2r_e^2+4M_ea^2r_e\left(1-y^2\right)-a^2\left(1-y^2\right)\tilde\epsilon_5(y)\right]}{4M_e^2r_{e}^2}d\psi^2\,.\label{3dembedingergosphere}
\end{align}

\begin{figure}
\begin{center}
\includegraphics[scale=0.45]{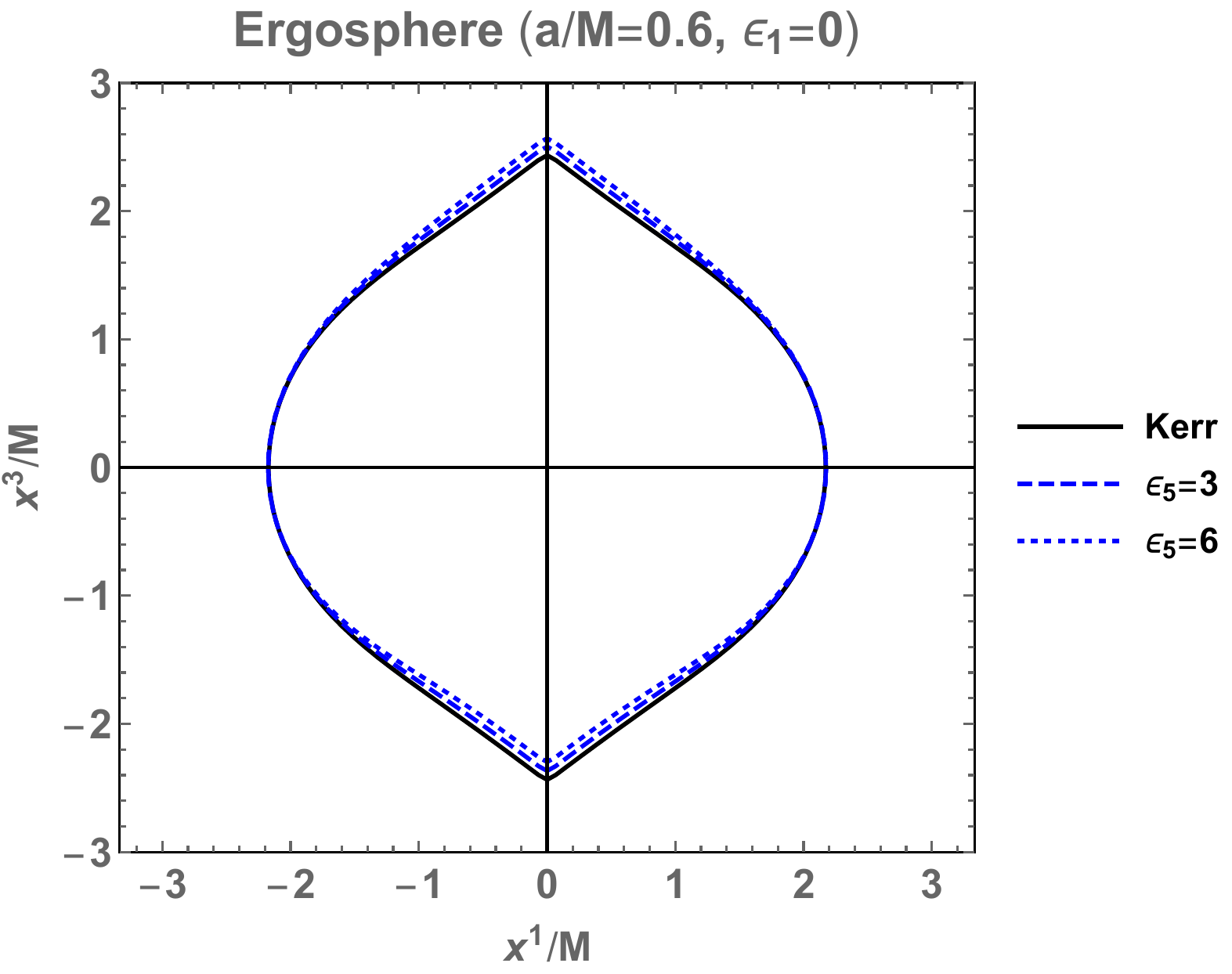}
\includegraphics[scale=0.45]{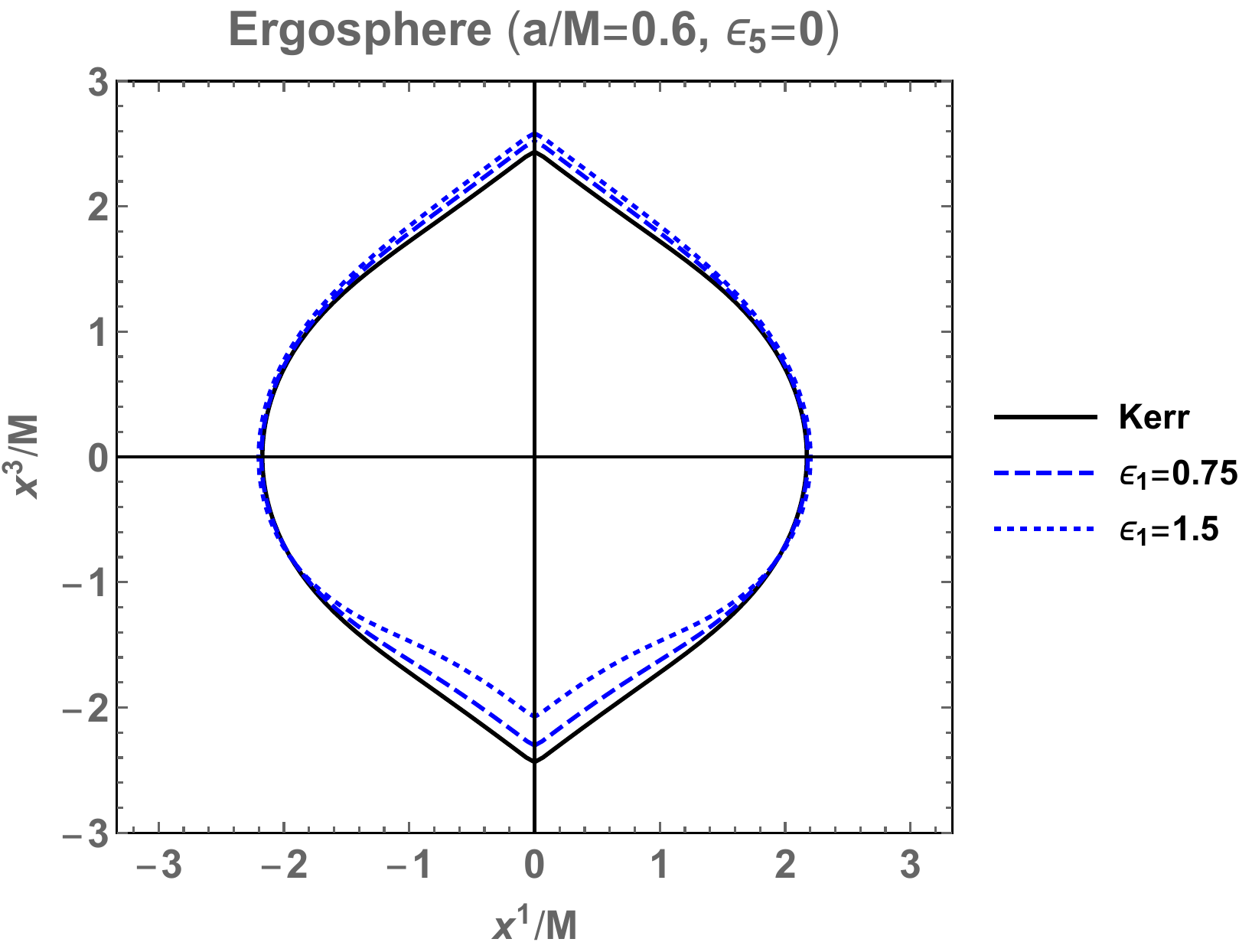}
\caption{\label{fig.embedego}The embeddings of the ergosurface in 3-dimensional Euclidean space. The black contour corresponds to the Kerr solution. In the left panel, we fix $\epsilon_1=0$ and $a/M=0.6$, then show the contour for different values of $\epsilon_5$. In the right panel, on the other hand, we fix $\epsilon_5=0$ and $a/M=0.6$, then show the contour for different values of $\epsilon_1$. In both cases, the $\mathbb{Z}_2$ asymmetry can be seen when changing $\epsilon_1$ or $\epsilon_5$.}
\end{center}
\end{figure}

As what we have done in the previous subsection, we use an isometric embedding to map the ergosurface into a 3-dimensional Euclidean space to visualize the ergosurface and the embedding shape of the ergosphere. This can be achieved by inserting the induced metric \eqref{3dembedingergosphere} into the metric \eqref{3dembeding}. In Figure~\ref{fig.embedego}, we assume $M(r)=M$, $\tilde\epsilon_1(y)=\epsilon_1M^2y$, and $\tilde\epsilon_5(y)=\epsilon_5M^2y$, and show the embedding of the ergosurface and the enclosed ergosphere in 3-dimensional Euclidean space. Similar to the case of event horizon in Figure~\ref{fig.embedeh}, the violation of $\mathbb{Z}_2$ symmetry can be seen from the ergosurface structure, when $\epsilon_1$ or $\epsilon_5$ is not zero.

\subsection{Ricci scalar}
\begin{figure}
\begin{center}
\includegraphics[scale=0.35]{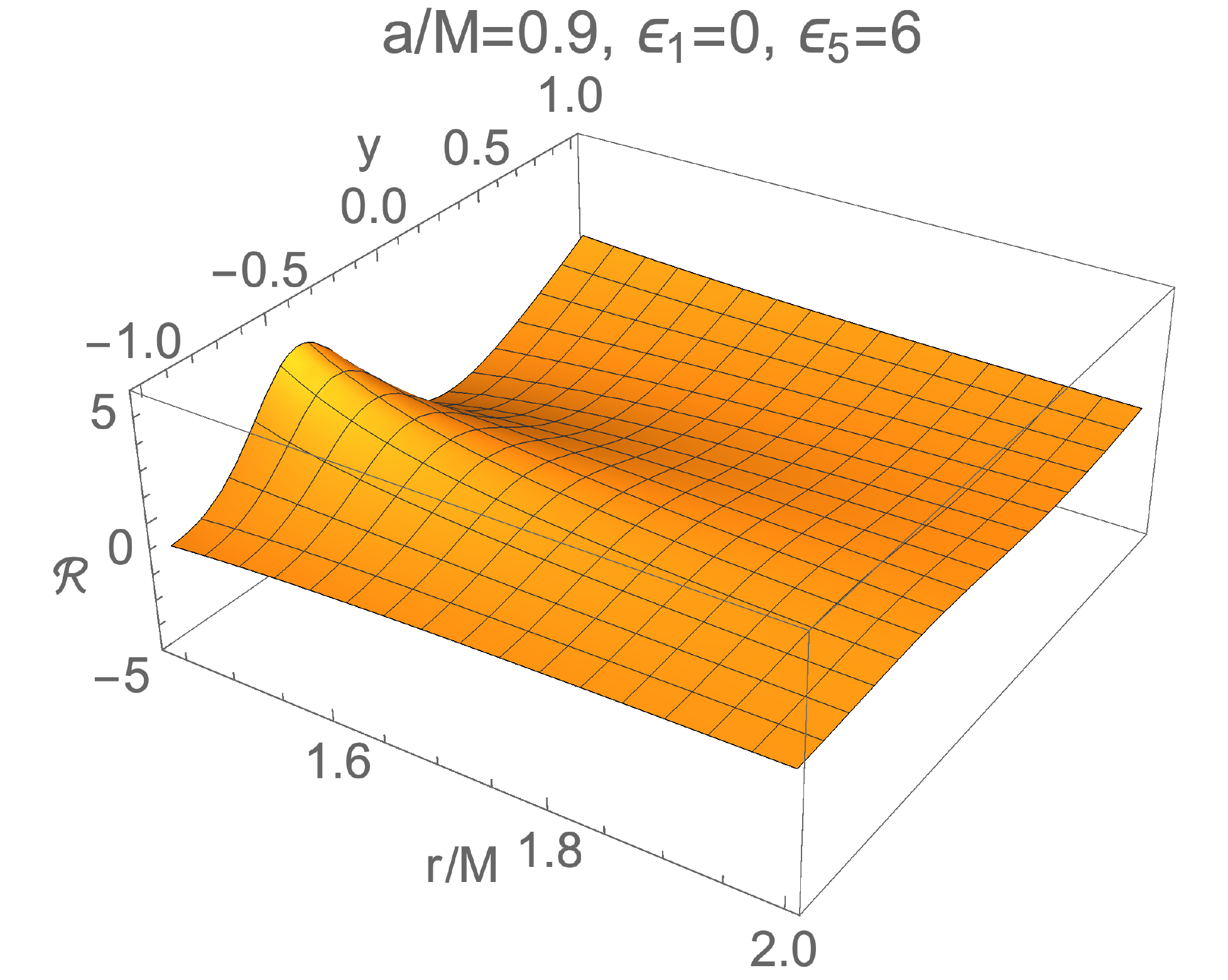}
\includegraphics[scale=0.35]{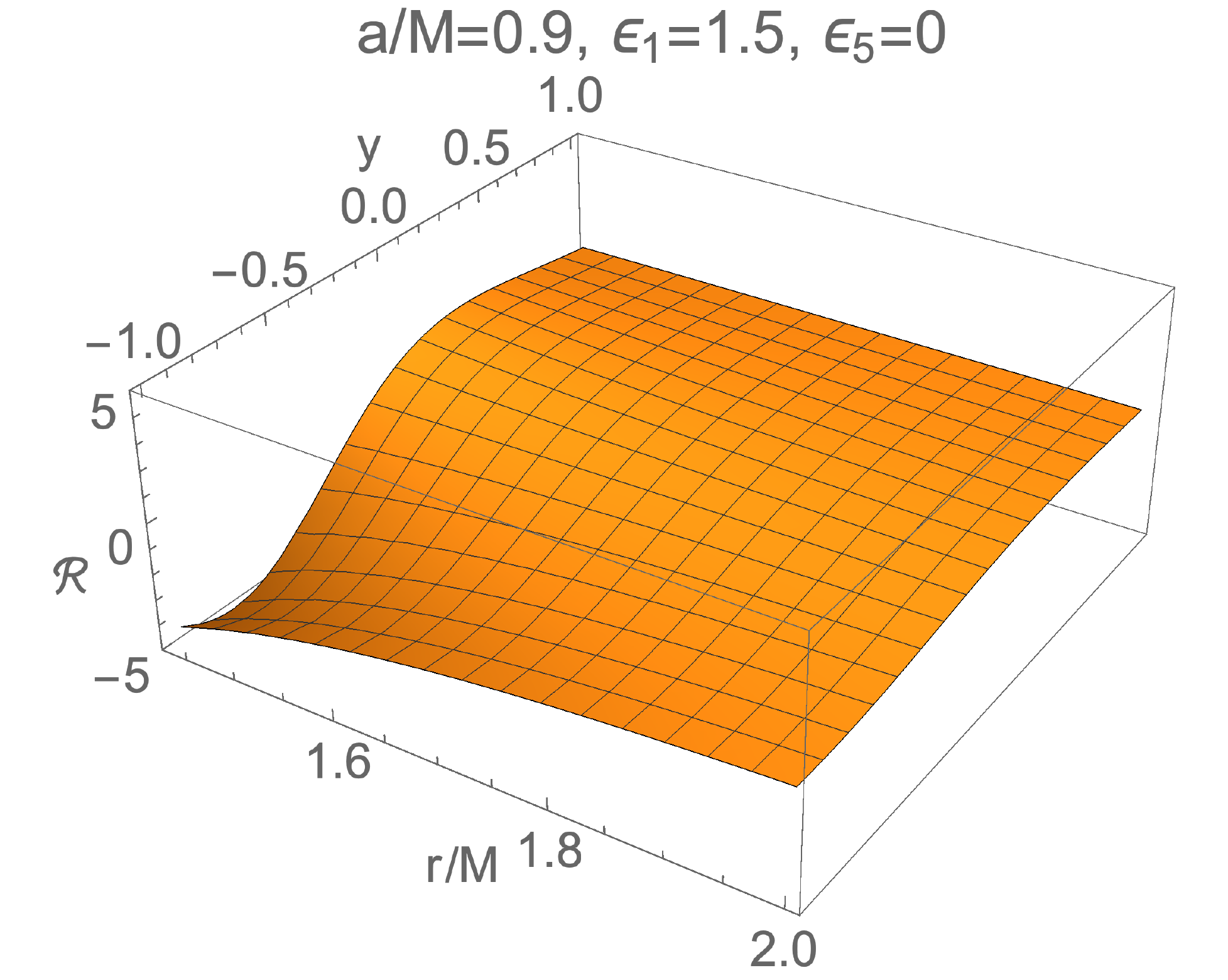}
\caption{\label{fig.Ricci}The Ricci scalar $\mathcal{R}(r,y)$ outside the horizon. In both panels, we assume $a/M=0.9$. In the left panel, we fix $\epsilon_1=0$ and $\epsilon_5=6$, while in the right panel, we assume $\epsilon_1=1.5$ and $\epsilon_5=0$.}
\end{center}
\end{figure}

The fact that the $\mathbb{Z}_2$ symmetry in the Kerr-like spacetime is broken can also be seen from the profile of the curvature invariant. The Ricci scalar $\mathcal{R}(r,y)$ of the Kerr-like spacetime can be approximated as (assuming $M(r)=M$):
\begin{align}
\mathcal{R}(r,y)\approx\frac{\left(y^2-1\right)\left(3\tilde\epsilon_{1}+\tilde\epsilon_5\right)_{,yy}+2y\left(3\tilde\epsilon_{1}+\tilde\epsilon_5\right)_{,y}-2\left(3\tilde\epsilon_{1}+\tilde\epsilon_5\right)}{r^4}\nonumber\\
+\frac{2M\left[12\tilde\epsilon_1+11\tilde\epsilon_5-2y\tilde\epsilon_{5,y}+\left(1-y^2\right)\tilde\epsilon_{5,yy}\right]}{r^5}+\mathcal{O}(r^{-6})\,,
\end{align}
when $r\rightarrow\infty$. In general, the Ricci scalar is not $\mathbb{Z}_2$ symmetric because of its arbitrary dependence on $y$. In Figure~\ref{fig.Ricci}, we show the Ricci scalar $\mathcal{R}(r,y)$ outside the horizon with respect to $r$ and $y$. In both panels, we fix the spin parameter to $a/M=0.9$ and assume that $M(r)=M$, $\tilde\epsilon_1(y)=\epsilon_1M^2y$, and $\tilde\epsilon_5(y)=\epsilon_5M^2y$. In the left panel, we fix $\epsilon_1=0$ and $\epsilon_5=6$, while in the right panel, we assume $\epsilon_1=1.5$ and $\epsilon_5=0$. It can be easily seen that the $\mathbb{Z}_2$ symmetry of the spacetime is broken by either changing $\epsilon_1$ or $\epsilon_5$.

\section{Shadows}\label{sec.shadow}
In the previous section, we have shown that the near-horizon structure as well as the ergosurface of the Kerr-like black hole is different from those of the Kerr black hole when the function $\tilde\epsilon_1(y)$ or $\tilde\epsilon_5(y)$ is non-zero. In the presence of these arbitrary functions, it is generically possible to break the $\mathbb{Z}_2$ symmetry of the spacetime, while the spacetime recovers the Kerr spacetime in the asymptotic region. It is natural to ask whether these deviations would leave some observational imprints, which enable us to test the Kerr-like spacetime under consideration. In this section, we will investigate the shadow of the Kerr-like black hole and see whether the deviation functions $\tilde\epsilon_1(y)$ and $\tilde\epsilon_5(y)$ would alter the shape or the size of the shadow contour. 

Essentially, the shadow of a rotating black hole is the impact parameter of the photon region around the black hole. This region consists of several spherical photon orbits, each with its own radius $r_p$, such that $R(r_p)=0$, $dR/dr|_{r_p}=0$, and $d^2R/dr^2|_{r_p}\ge0$. Considering our Kerr-like spacetime metric, Eqs.~\eqref{RE} and \eqref{ThetaE} can be rewritten as
\begin{align}
\frac{R(r)}{E^2}&=\left(r^2+a^2-a\xi\right)^2-\Delta\left[\eta+\left(\xi-a\right)^2\right]\,,\\
\frac{\Theta(\theta)}{E^2}&=\left[\eta+\cos^2\theta\left(a^2-\xi^2\csc^2\theta\right)-\tilde\epsilon_5(y)\right]\,.
\end{align}
The equations $R(r_p)=0$ and $dR/dr|_{r_p}=0$ give
\begin{align}
\xi(r_p)=\frac{1}{a}\left(r^2+a^2-\frac{4\Delta r}{\Delta'}\right)\Bigg|_{r=r_p}\,,\label{xietaphoton1}\\
\eta(r_p)=\frac{16r^2\Delta }{\Delta'^2}\Bigg|_{r=r_p}-\left[\xi(r_p)-a\right]^2\,,\label{xietaphoton2}
\end{align}
where the prime stands for the derivative with respect to $r$. Note that $\Theta$ should not be negative for a viable photon trajectory. Given a spherical photon orbit with a radius $r_p$, the azimuthal angular momentum $\xi$ and the Carter-like constant $\eta$ of a photon moving on this orbit are determined by Eqs.~\eqref{xietaphoton1} and \eqref{xietaphoton2}, respectively. It can be shown that the azimuthal angular momentum $\xi$ and the orbital radius $r_p$ have an one-to-one correspondence. In addition, along each spherical photon orbit, the latitude $\theta$ oscillates between its own extreme values. Given a set of $\xi$ and $\eta$, these extreme latitudes can be obtained by solving the equation $\dot\theta=0$. In the absence of $\tilde\epsilon_5(y)$, the motion on each spherical orbit is $\mathbb{Z}_2$ symmetric because $\Theta(y)$ contains only even functions of $y$. However, if $\tilde\epsilon_5(y)$ is not zero, the $\mathbb{Z}_2$ symmetry could be broken and the extreme latitudes on the north and the south hemispheres could be different. Note that only the photons with a zero azimuthal angular momentum, that is, $\xi=0$, are able to reach the axis of symmetry ($y=\pm1$). For a discussion of the spherical photon orbits around a Kerr black hole, we refer the reader to the paper \cite{Teo:2003} and the review \cite{Perlick:2010zh}.

\begin{figure}
\begin{center}
\includegraphics[scale=0.38]{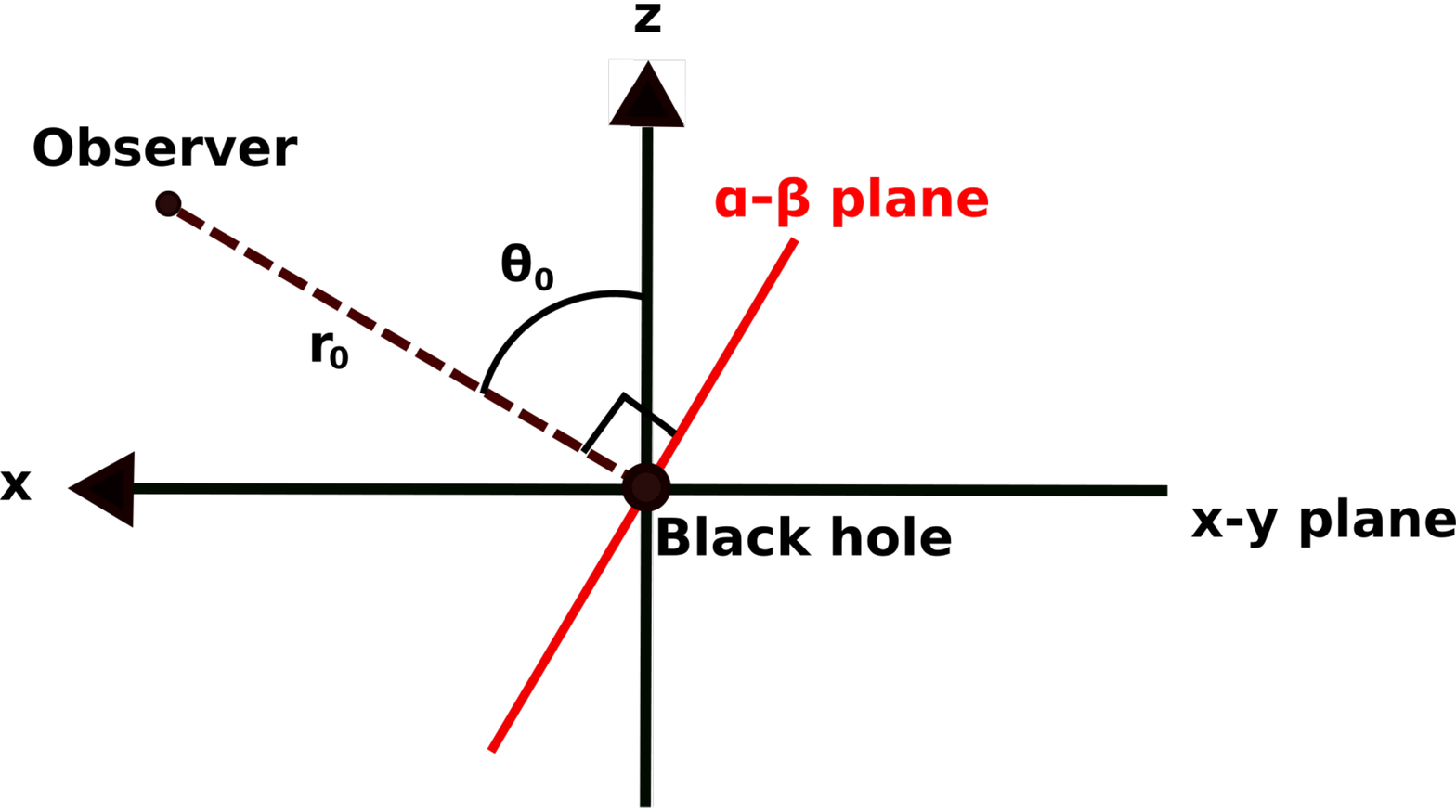}
\caption{\label{cele1}The schematic plot of the celestial coordinates. The black hole is at the origin and the observer is on the $x$-$z$ plane. The $z$ axis is assumed to be the axis of symmetry of the black hole. The position of the observer can be represented with spherical coordinates ($r_0,\theta_0,\psi_0$) and one can choose $\psi_0=0$. As seen from the observer, the black hole shadow is projected on the $\alpha$-$\beta$ plane.}
\end{center}
\end{figure}

In order to visualize the apparent shape of a shadow, we adopt the celestial coordinates ($\alpha,\beta$) which lie on the celestial plane of the observer. The illustration is depicted in Figure~\ref{cele1}. The coordinate $\alpha$ is the apparent perpendicular distance between the edge of the shadow and the axis of symmetry ($z$-axis). The coordinate $\beta$, on the other hand, is the apparent perpendicular distance between the edge of the shadow and the $y$-axis. In an asymptotically flat spacetime, the celestial coordinates can be expressed as \cite{Vazquez:2003zm}
\begin{equation}
\alpha=\lim_{r_0\rightarrow\infty}\left(-r_0^2\sin\theta_0\frac{d\psi}{dr}\bigg|_{r_0,\theta_0}\right)\,,\qquad \beta=\lim_{r_0\rightarrow\infty}\left(r_0^2\frac{d\theta}{dr}\bigg|_{r_0,\theta_0}\right)\,,\label{alphabetadef}
\end{equation} 
where $r_0$ is the distance between the observer and the black hole, and $\theta_0$ is the inclination angle between the rotation axis ($z$-axis in Figure~\ref{cele1}) of the black hole and the direction to the observer. Finally, using the geodesic equations \eqref{dotr}, \eqref{dottheta}, \eqref{dott3}, and \eqref{dotpsi3}, one can obtain the expressions of the celestial coordinates for the Kerr-like black hole as follows:
\begin{equation}
\alpha=-\frac{\xi}{\sin\theta_0}\,,\qquad\beta=\pm\sqrt{\eta+a^2\cos^2\theta_0-\xi^2\cot^2\theta_0-\tilde\epsilon_5(y_0)}\,,\label{alphabetarelation}
\end{equation}
where $y_0\equiv\cos\theta_0$. Combining Eq.~\eqref{alphabetarelation} and the parametrizations $\xi(r_p)$ and $\eta(r_p)$, the contour of the shadow can be obtained by parametrizing $\alpha$ and $\beta$ with the running variable $r_p$.

It can be seen that the shadow contour is completely blind to the deviation function $\tilde\epsilon_1(y)$. Also, even though the spacetime structure could violate $\mathbb{Z}_2$ symmetry, the shadow contour is symmetric with respect to the horizontal axis ($\alpha$-axis). This fact has been pointed out in Refs.~\cite{Cunha:2018uzc} and \cite{Grenzebach:2014fha}, in which the latter discusses the shadow of the Kerr-Newman-NUT black hole with a cosmological constant. The reason for the preservation of this symmetry in the shadow contour is related to the separability of the geodesic equations. More precisely, the separation of the Hamilton-Jacobi equation implies that in the $\theta$ sector of the geodesic equation, the corresponding component of the 4-momentum $p_\theta$ always appears in the form of $p_\theta^2$ \cite{Cunha:2018uzc}. Since the coordinate $\beta$ is proportional to $p_\theta$ \cite{1973ApJ...183..237C} (see Eq.~\eqref{alphabetadef}), the shadow contour is symmetric with respect to the horizontal axis. 

\begin{figure}
\begin{center}
\includegraphics[scale=0.5]{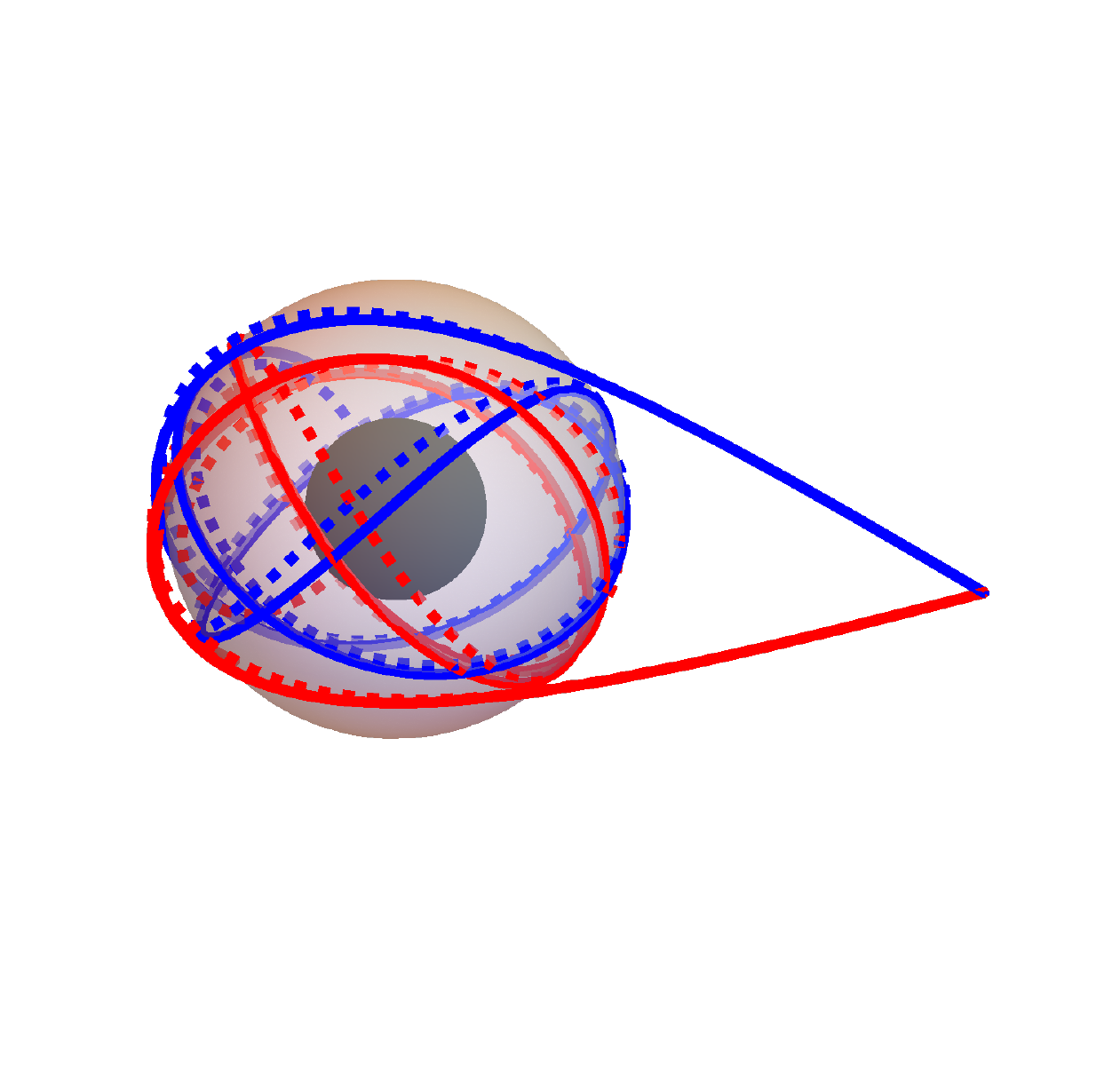}
\caption{\label{fig.ray}The trajectories of two light rays contributing to two points nearly on the shadow contour, with celestial coordinates $(\alpha_p,\pm\beta_p)$. The observer is on the right-hand side where light rays converge. The solid curves correspond to the result of the Kerr-like black hole with $\epsilon_5=6$, while the dashed curves correspond to the trajectories in the Kerr spacetime. Photons following all these trajectories would approach to the same photon sphere (the outer sphere) and undergo spherical motions. The event horizon is illustrated by the inner black sphere.}
\end{center}
\end{figure}

In Figure~\ref{fig.ray}, we assume $M(r)=M$ and $\tilde\epsilon_5(y)=\epsilon_5M^2y$, and show the photon trajectories contributing to two points nearly on the shadow contour, with celestial coordinates $(\alpha_p,\pm\beta_p)$. The observer is on the right-hand side where light rays converge. The solid curves are the result in the Kerr-like spacetime with $\epsilon_5=6$, while the dashed curves represent that in the Kerr spacetime. It can be shown that photons following all these curves would approach the same photon sphere (the outer sphere) and undergo spherical motions. This explains why the shadow contour would appear symmetric with respect to the horizontal axis. It should be noticed that, although the blue and the red curves correspond to the same photon sphere even if $\epsilon_5\ne0$, the spherical motions of these two curves are $\mathbb{Z}_2$ asymmetric. One can indeed see that the solid curve and the dashed curve are distinguishable on the photon sphere, which the latter is $\mathbb{Z}_2$ symmetric.

It should be stressed that in this paper, we construct the Kerr-like spacetime by using the general PK metric. We require the spacetime to be asymptotically flat and focus on the deviation functions $\tilde\epsilon_i(y)$ which could generically break the $\mathbb{Z}_2$ symmetry of the spacetime. The $\mathbb{Z}_2$ symmetry of the shadow contour turns out to be a common property shared by the family of such general Kerr-like spacetimes.

\begin{figure}
\begin{center}
\includegraphics[scale=0.45]{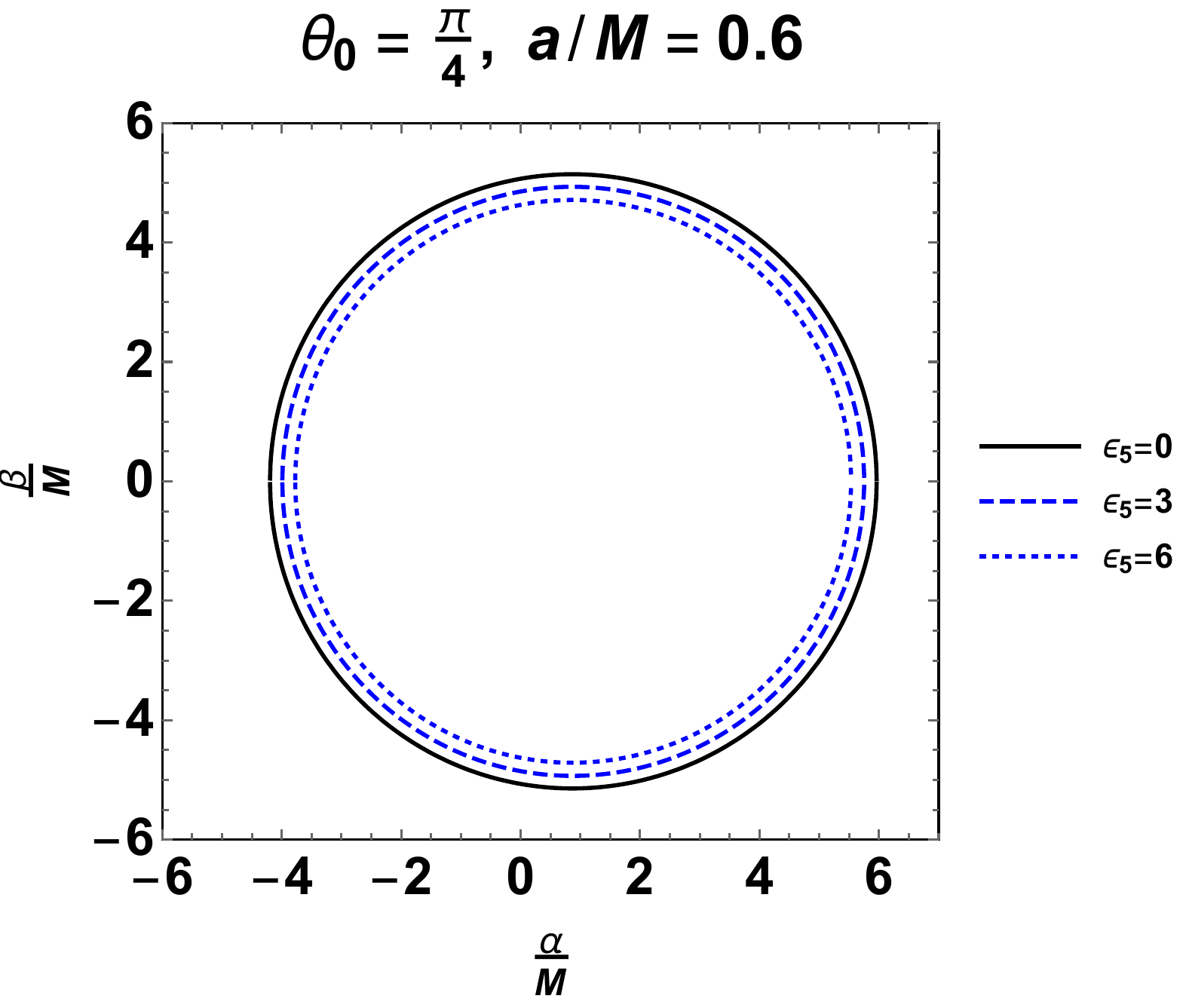}
\includegraphics[scale=0.45]{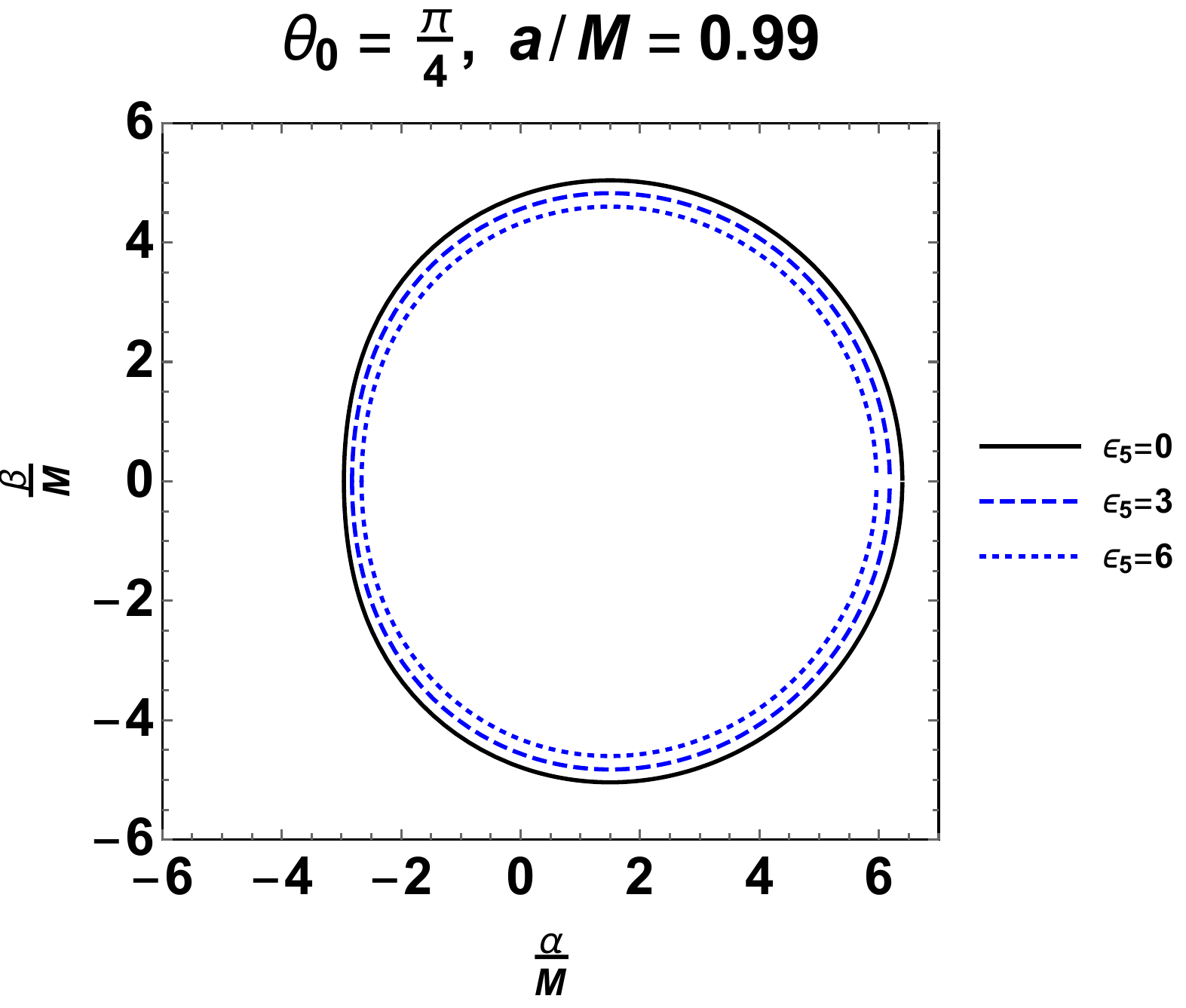}
\caption{\label{fig.shadow}The shadow contours for different values of $\epsilon_5$ are shown. Note that the shadow contour is completely blind to $\tilde\epsilon_1(y)$. We have fixed the inclination angle $\theta_0=\pi/4$. In the left panel, we fix $a/M=0.6$, while in the right panel we choose $a/M=0.99$.}
\end{center}
\end{figure}

In Figure~\ref{fig.shadow}, we show the shadow contours of the Kerr-like black hole for different values of $\epsilon_5$. Note that the shadow contour is completely blind to the deviation function $\tilde\epsilon_1(y)$. We fix the inclination angle $\theta_0=\pi/4$ and assume the spin to be $a/M=0.6$ in the left panel. In the right panel, we show the contours for $a/M=0.99$ (the extremal case corresponds to $a/M=1$). One can see that the contour is more distorted when the spin increases, as expected. According to Figure~\ref{fig.shadow}, it can be directly seen that the apparent size of the shadow contour shrinks when $\epsilon_5$ increases, for the current choice of $\theta_0$. However, it seems that the change of $\epsilon_5$ only affects the distortion of the contour by a very small amount.

In order to investigate whether it is possible to extract the information about black hole parameters from shadow contours, we consider the observables that characterize the shadow contour. We use the method developed in Ref.~\cite{Hioki:2009na} in which the authors defined two parameters $R_S$ and $D_S$. The former corresponds to the apparent size of the shadow, while the latter quantifies its distortion in shape. There are also several possible observables one can define from a shadow contour \cite{Ghasemi-Nodehi:2015raa,Abdujabbarov:2015xqa,Tsukamoto:2014tja,Kumar:2018ple}. In this paper, we consider the simplest but very seminal one proposed in Ref.~\cite{Hioki:2009na}. 

\begin{figure}
\begin{center}
\includegraphics[scale=0.38]{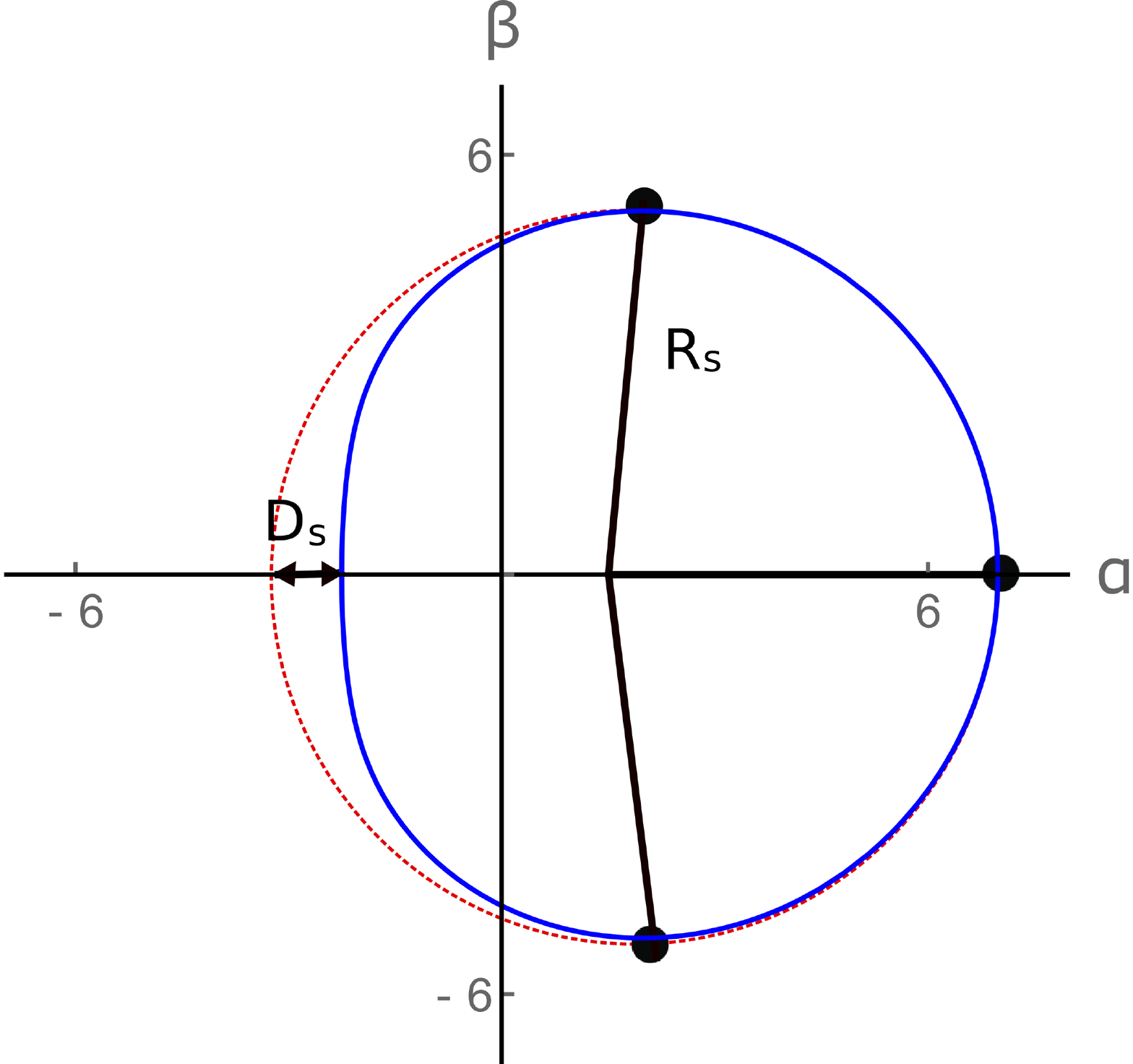}
\caption{\label{fig.shadowpa}This figure illustrates schematically the definition and the geometrical meaning of the observables $R_S$ and $D_S$.}
\end{center}
\end{figure}

The schematic plot and the geometrical meaning of the observables $R_S$ and $D_S$ \cite{Hioki:2009na} are illustrated in Figure~\ref{fig.shadowpa}. In this figure, the black hole shadow is depicted by the blue contour. The apparent radius $R_S$ of a shadow contour is defined by considering a reference circle (the red dotted circle) passing through the top, bottom, and the rightmost points on the shadow. This circle is uniquely defined by these three points and we define its radius as the apparent size of the shadow $R_S$. On the other hand, the distortion parameter $D_S$ is defined by the apparent distance between the leftmost point of the reference circle and that of the shadow contour. The distortion parameter thus measures the amount of the shadow contour deviates from a perfect circle. It is also common to define a dimensionless parameter $D_S/R_S$ to quantify the distortion of the black hole shadow \cite{Hioki:2009na}. 

\begin{figure}
\begin{center}
\includegraphics[scale=0.45]{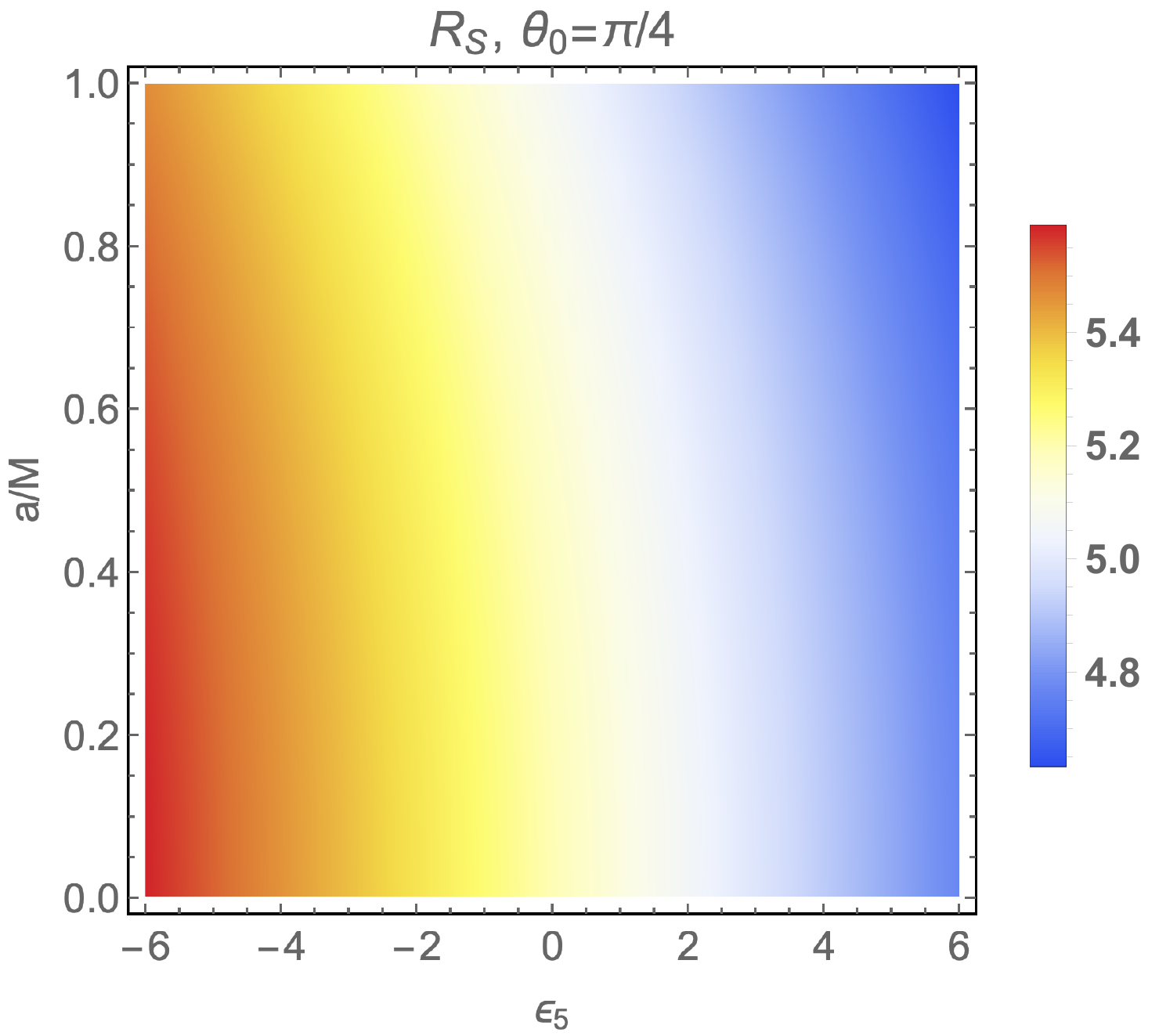}
\includegraphics[scale=0.45]{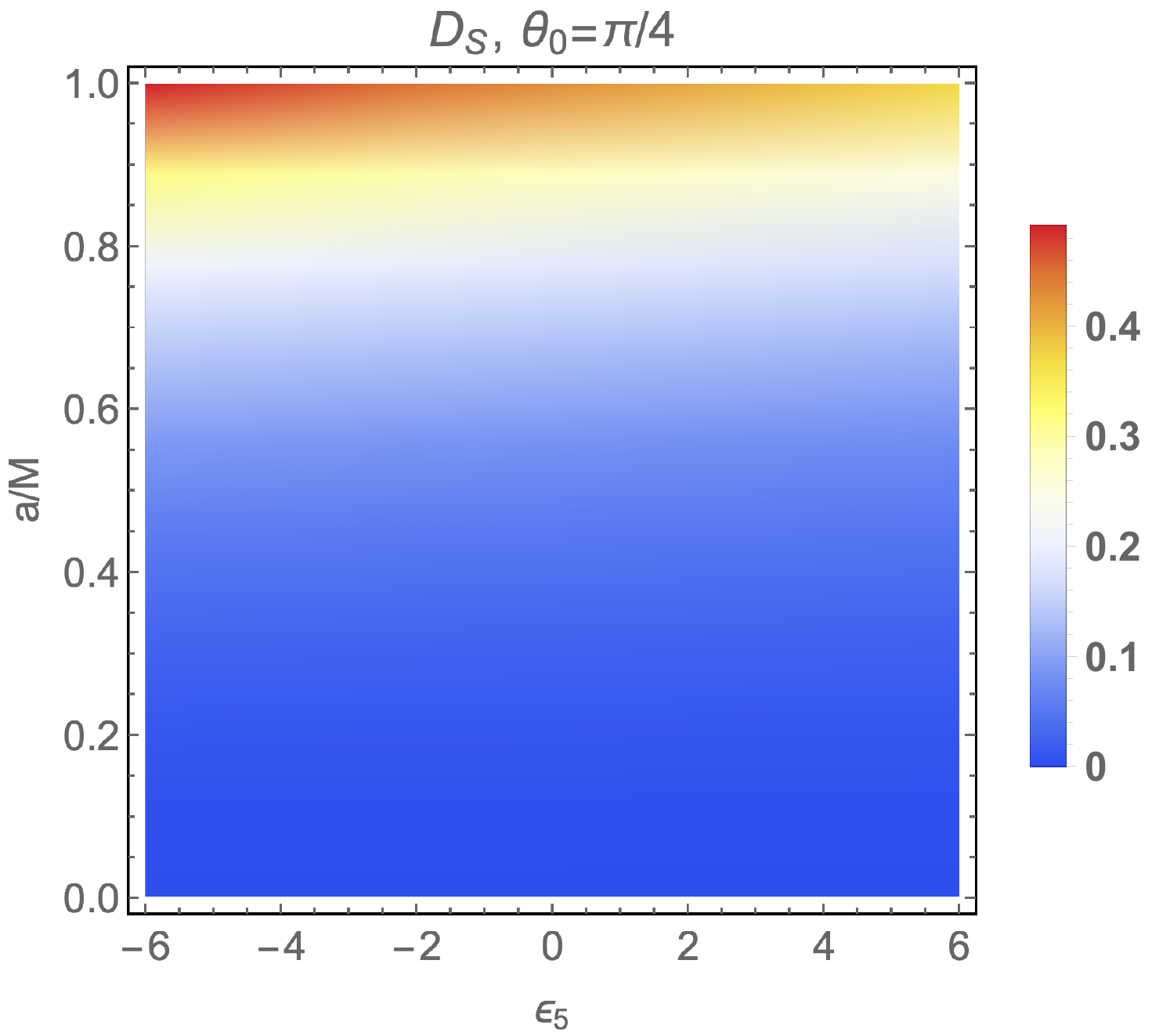}
\includegraphics[scale=0.45]{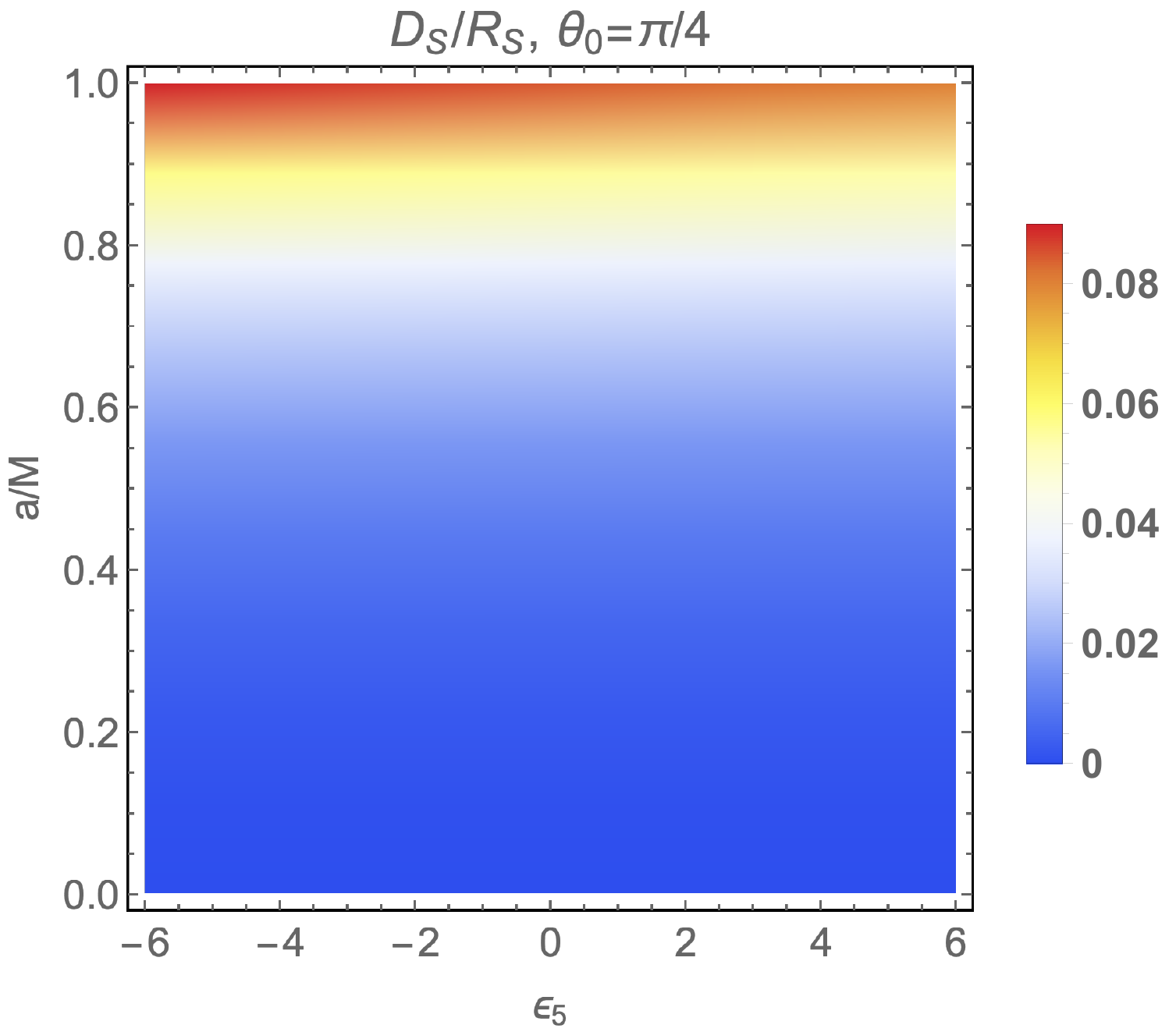}
\caption{\label{fig.a5}These density profiles show how the observables $R_S$ (top-left), $D_S$ (top-right), and $D_S/R_S$ (bottom) are changed with respect to $\epsilon_5$ and the spin parameter. Here we have fixed the inclination angle $\theta_0=\pi/4$.}
\end{center}
\end{figure}

\begin{figure}
\begin{center}
\includegraphics[scale=0.45]{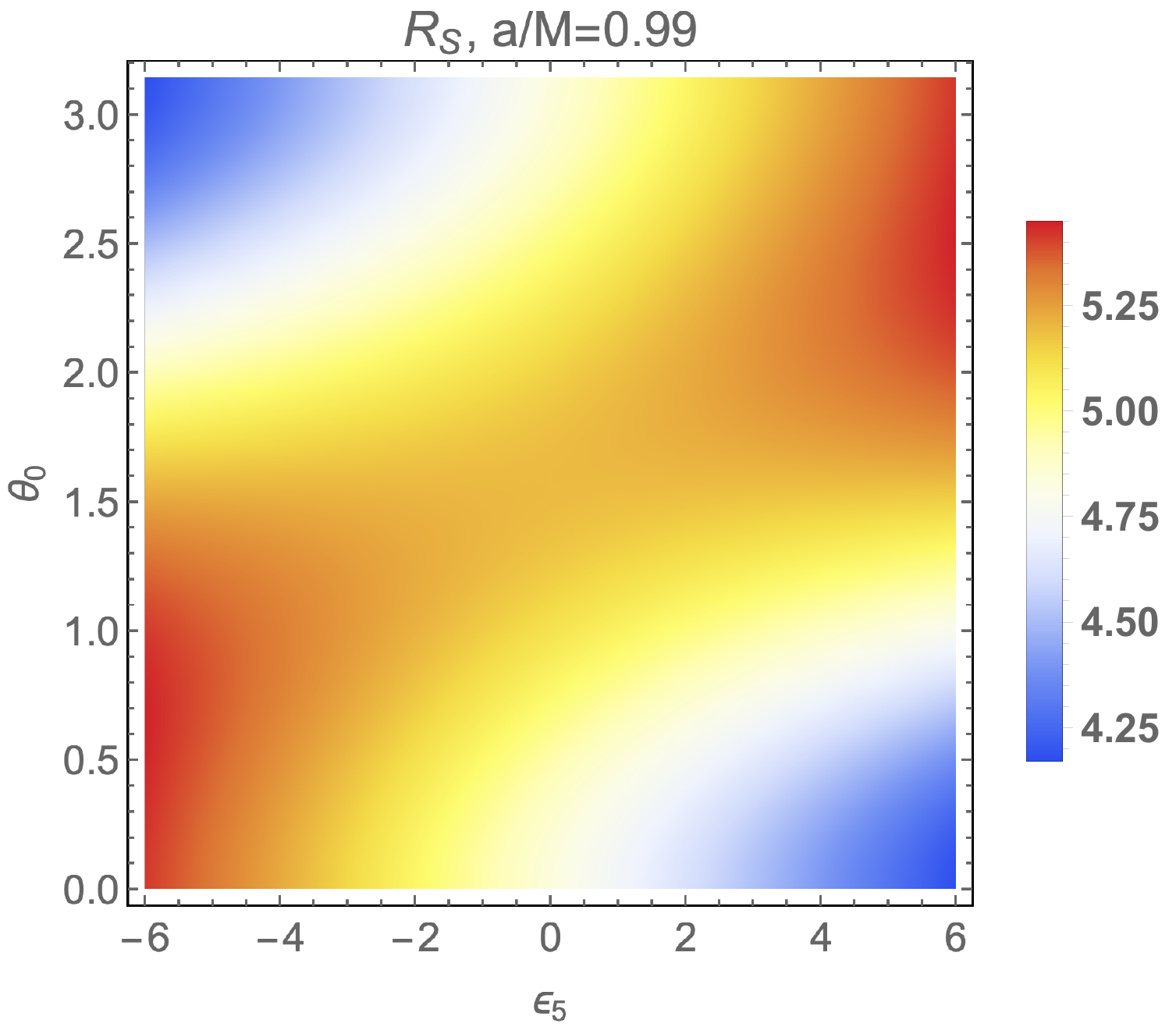}
\includegraphics[scale=0.45]{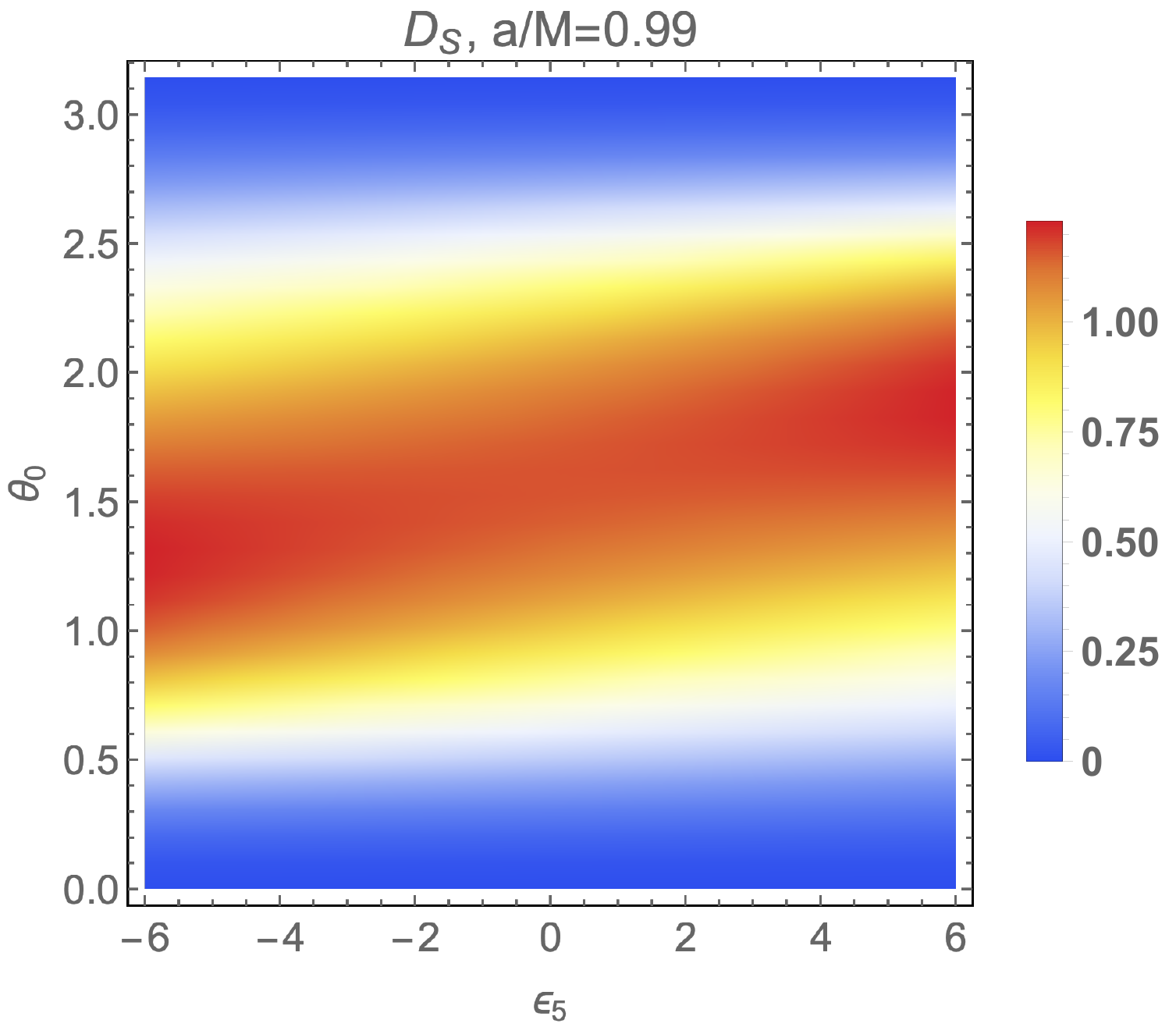}
\includegraphics[scale=0.45]{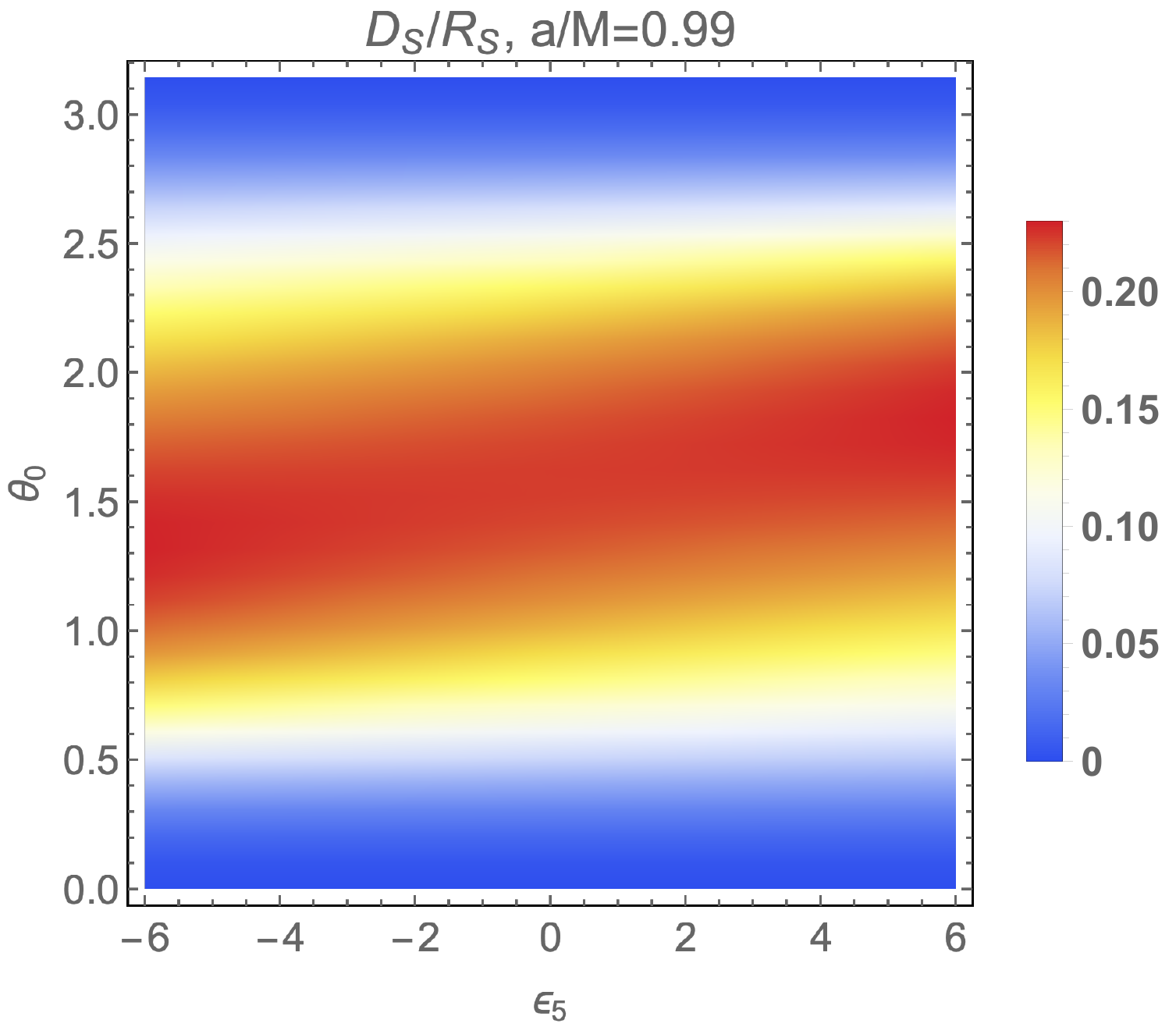}
\caption{\label{fig.theta5}These density profiles show how the observables $R_S$ (top-left), $D_S$ (top-right), and $D_S/R_S$ (bottom) are changed with respect to $\epsilon_5$ and the inclination angle. Here we have fixed the spin parameter $a/M=0.99$.}
\end{center}
\end{figure}

In Figure~\ref{fig.a5}, we assume $M(r)=M$, $\tilde\epsilon_5(y)=\epsilon_5M^2y$, and $\theta_0=\pi/4$, then show how the observables $R_S$ (top-left), $D_S$ (top-right), and $D_S/R_S$ (bottom) change with respect to the spin parameter $a$ and $\epsilon_5$. As one can see from the top-left panel, the apparent size of the shadow for a given $\epsilon_5$ would shrink a little bit when the black hole is getting more extremal. Furthermore, it can be seen that increasing $\epsilon_5$ would reduce the apparent size of the shadow when $0\le\theta_0\le\pi/2$ (see also the top-left panel of Figure~\ref{fig.theta5}). As for the distortion in shape, one can infer from the top-right and the bottom panels that changing $\epsilon_5$ has almost no contribution to the distortion parameter $D_S$. When fixing the inclination angle $\theta_0$ and assuming a constant mass function, the distortion in shape is mostly determined by the spin parameter. 

In Figure~\ref{fig.theta5}, we assume a nearly extremal black hole $a/M=0.99$ and show how the observables change with respect to $\epsilon_5$ and the inclination angle $\theta_0$. According to the top-left panel, it seems that the change of the apparent size with respect to changing the parameters $(\theta_0,\epsilon_5)$ is non-trivial. In fact, when $\theta_0=\pi/2$, the apparent size $R_S$ remains unaltered when changing $\epsilon_5$. This is expected because $\epsilon_5$ appears in the expression of the celestial coordinates \eqref{alphabetarelation} only in the form of $\epsilon_5\cos\theta_0$. In fact, if $\tilde\epsilon_5|_{y=0}=0$, the edge-on Kerr-like shadow is indistinguishable from its Kerr counterpart. As for the distortion parameter, it can be seen from the top-right and bottom panels that the distortion parameter $D_S$ is very insensitive to the change of $\epsilon_5$.

\section{Conclusion}\label{sec.conclu}
In this paper, we study the physical properties and the shadow contour of a class of Kerr-like black holes, whose $\mathbb{Z}_2$ symmetry is generically broken. Such black holes could arise in some candidates of effective theories of quantum gravity, especially when parity-violating terms appear in the higher-derivative interactions. In this work, we adopt an theory-agnostic approach and construct a class of Kerr-like spacetimes in which the $\mathbb{Z}_2$ asymmetry is quantified by two arbitrary functions of polar angle: $\tilde\epsilon_1$ and $\tilde\epsilon_5$. The metric, although not a solution of any particular existing theory of gravity, is able to parametrize the deviations from Kerr black hole of our interest, e.g. the amount of $\mathbb{Z}_2$ symmetry violation. In addition, the Kerr-like metric under consideration possesses the following crucial properties: the asymptotic flatness and the existence of a Carter-like constant. The latter property implies that the geodesic equations are completely separable and they can be written in first-order form. 

After constructing the Kerr-like metric, we identify the locations of the event horizon and the ergosurface, respectively. Then, we exhibit the $\mathbb{Z}_2$ asymmetry of the spacetime by using an isometric embedding to map the induced metrics on the event horizon and the ergosurface into a 3-dimensional Euclidean space. We show that by changing $\tilde\epsilon_1$ or $\tilde\epsilon_5$ from zero, the $\mathbb{Z}_2$ symmetry of the spacetime could be broken. The $\mathbb{Z}_2$ asymmetry can also be seen from the behavior of the Ricci scalar, as has been depicted in Figure~\ref{fig.Ricci}. 

Furthermore, we study the shadow contour of this Kerr-like black hole. It turns out that the shadow contour is completely blind to the deviation function $\tilde\epsilon_1$. On the other hand, the other deviation function, $\tilde\epsilon_5$, does alter the shadow contour. By assuming $\tilde\epsilon_5=\epsilon_5M^2\cos\theta$, we find that for a given spin parameter, increasing the value of $\epsilon_5$ would shrink (expand) the apparent size of the shadow, if the inclination angle $\theta_0$ is smaller (larger) than $\pi/2$. However, changing the parameter $\epsilon_5$ seems to hardly affect the distortion of the shadow contour. For this type of Kerr-like black holes, the distortion of the shadow contour, that is, $D_s/R_s$, is mostly determined by the spin and the inclination angle. This essentially means that the deviation function $\tilde\epsilon_5$ can only be tested when the distance to the black hole and the black hole mass can be measured with great precision. 

Another important result is that even though the $\mathbb{Z}_2$ symmetry of the spacetime is generically broken, the shadow contour is still symmetric with respect to the horizontal axis, irrespective of the inclination angle. This discovery has been pointed in Refs.~\cite{Grenzebach:2014fha} and \cite{Cunha:2018uzc}, respectively for the Kerr-Newman-NUT black hole with a cosmological constant, and for a particular black hole spacetime generated by extra fields non-minimally coupled to gravity. The reason is associated to the separability of the Hamilton-Jacobi equation for geodesic equations. The vertical angular distance of a point on the shadow contour to the horizontal axis is proportional to $d\theta/dr$ on the light curve, evaluated at the location of the observer. According to the geodesic equation \eqref{dottheta}, upon reaching the observer, photons with a positive and a negative $d\theta/dr$ possess identical conserved quantities $\eta$ and $\xi$. Therefore, they would be mapped onto the same photon sphere around the black hole (see Figure \ref{fig.ray}). That is why the shadow contour is always symmetric with respect to the horizontal axis. In Ref.~\cite{Grenzebach:2014fha}, the author has shown that the shadow contour is symmetric even if the observer is located at a finite distance $r_0$ from the Kerr-Newman-NUT black hole. In this paper, we have shown that it is also true for a general Kerr-like black hole. Essentially, this result provides another evidence that the shadow of a black hole is not really sensitive to the intrinsic geometric structure of the event horizon \cite{Cunha:2018gql}.

Given that it is challenging to test $\tilde\epsilon_1$ and $\tilde\epsilon_5$ with shadow contours, one could resort to black hole spectroscopy by considering perturbations of this type of black holes. The deviation functions may leave imprints on the quasinormal mode frequencies of the black hole. Another possible extension is to relax the assumption that a Carter-like constant exists. This would increase the diversity of viable models and the geodesic equations would not be separable anymore. It would be interesting to see how the shadow of such black holes acquires its novel characteristics, such as the asymmetry with respect to the horizontal axis. We will leave these issues for future works.

\acknowledgments
CYC would like to express his gratitude to Jiunn-Wei Chen for fruitful suggestions and discussions on this work. CYC is supported by Ministry of Science and Technology (MOST), Taiwan, through No. 107-2119-M-002-005
and No. 108-2811-M-002-682. He is also supported by Leung Center for Cosmology and Particle Astrophysics
(LeCosPA) of National Taiwan University, and Taiwan National Center for Theoretical Sciences (NCTS).

\end{document}